\documentstyle[eqsecnum,aps,epsf]{revtex}
\begin{document}
\draft
\title{ The Extended Coupled Cluster Treatment of \\ 
Correlations in Quantum Magnets}
\author{ J. Rosenfeld, N. E. Ligterink, and 
R. F. Bishop\footnote{e-mail: R.F.Bishop@umist.ac.uk}}
\address{Department of Physics, University of Manchester 
Institute of Science and Technology (UMIST), P.O. Box 88, \\ 
Manchester, M60 1QD, UK} 
\date{\today}
\maketitle
\begin{abstract}The spin-half
{\it XXZ} model on the linear chain and the square lattice 
are examined with the
 extended coupled cluster method (ECCM) 
of quantum many-body theory.
We are able to describe both the Ising-Heisenberg phase and 
the {\it XY}-Heisenberg phase, starting 
from known wave functions in the Ising limit 
and at the phase transition point between the {\it XY}-Heisenberg and 
ferromagnetic phases, respectively, and 
by systematically incorporating correlations on top of 
them. 
The ECCM yields good numerical results via a diagrammatic approach, 
which makes the numerical implementation of higher-order truncation schemes 
feasible.
In particular, the best non-extrapolated coupled cluster result for the 
sublattice magnetization is obtained, which indicates the employment of an improved wave function.
Furthermore, the ECCM finds the expected qualitatively different behaviours 
of the linear chain and the square lattice cases. 
\end{abstract}

\section{Introduction}
The extended coupled cluster method (ECCM) \cite{{arp:rpo},{arp:paj},{rb1:ja},{rb2:ja},{rb3:ja}} has not previously been applied to lattice spin systems,
unlike the normal coupled cluster method (NCCM) 
\cite{{rb1:ja},{rb2:ja},{rb3:ja},{coe:fc},{ci:ciz},{ku:luh},{rfb:luh},{bi:kum},{br:be},{bi:rbi},{bi:ss}}, 
which is a restricted version of the ECCM at a given level of approximation, and which has been widely 
implemented for these systems.
 The primary aim of this paper 
is to apply the ECCM to the spin-half anisotropic Heisenberg (or 
{\it XXZ}) model, in 
order to obtain numerical results for the ground-state energy 
and the sublattice magnetization, 
and thereby to investigate the usefulness of the method in the study of 
quantum phase transitions. The ECCM and NCCM are 
{\it ab initio} techniques of microscopic quantum many-body theory, 
generically known as the coupled cluster method (CCM). 
The ECCM, by contrast to the NCCM, 
completely characterizes a system 
in terms of a set of basic amplitudes, all of which are
 linked-cluster quantities.

The two-dimensional spin-half {\it XXZ} model is expected to have a 
second-order phase transition at the isotropic point, $\Delta =1$, where the 
system is in a unique critical phase. The transition is expected
 to be accompanied by some change of symmetry of the ground-state 
wave function. In particular, 
Laughlin \cite{la:lau} speculates that the physics of the 
isotropic
 point can be understood 
as a gauge theory with massless excitations.
In general, theoretical study of the isotropic point
 requires some prior knowledge of the phases on either
 side of the
 isotropic point.
In practice, the ordering and symmetry of the known wave functions
in these phases influence the predicted state at the isotropic point.
Therefore, the second aim of this paper is to examine how the 
CCM, in particular, is affected by this universal 
 problem in the study of phase transitions.

The general {\it XXZ} model 
Hamiltonian has $Z(2)\otimes U(1)_{xy}$ symmetry, except 
at the isotropic 
point where it has $SU(2)$ symmetry.
For the square-lattice case the ground-state wave function is 
expected to yield broken $U(1)_{xy}$ symmetry in the {\it XY}-like region, 
and there is broken $Z(2)$ symmetry in the Ising limit. 
Approaching the isotropic point from either side with the 
CCM yields different 
ground-state wave functions at the isotropic point itself, 
which is due to the special nature of the isotropic point.

 Any CCM calculation on spin systems 
involves adding correlations between spins, on top of 
those already contained in a separately 
chosen model or reference state, in order to 
produce the true quantum-mechanical ground-state wave function.
Therefore, the important physical characteristics 
of a system are incorporated into the 
CCM by the choice of the model state and by the inclusion of 
particular configurations for the correlations.
Hence, we can now refine and reformulate our second aim to be an examination 
of the effect of the choice of 
model state. In particular, we shall be interested 
in the interplay between the symmetry 
of the model state and the 
symmetry of the Hamiltonian influencing the ground-state wave function.
This is exemplified by the CCM treatment of the linear-chain 
case, which yields 
artificially broken $U(1)_{xy}$ and $SU(2)$ symmetry. 
This can only be due to the choice of model state, since exact results 
show that the symmetry of the 
{\it XXZ} model Hamiltonian is not broken in the {\it XY}-like
region and at the isotropic point.

The primary aim of this paper 
is motivated by numerical evidence, 
from a number of techniques, 
of long-range order (LRO) at the isotropic point for the 
square-lattice case.
Numerical results from techniques
such as spin-wave theory \cite{{an:and},{kan:yo},{og:og}}, 
high-accuracy quantum
 Monte Carlo (QMC) simulations \cite{kr:kru},
 and series expansions\cite{{si:sin},{zh:wh}}, 
yield a sublattice magnetization at the isotropic point
of approximately $61$--$62\%$ of the classical value 
arising from perfect N\'{e}el ordering. However, it is still possible that 
each of these techinques may underestimate the quantum fluctuations in a 
similar fashion, since they all start from an ordered state\cite{wa:go}.

Rigorous results from the Bethe 
ansatz\cite{{be:bet},{Hu:hul},{or:ya}} for the spin-half anisotropic 
Heisenberg 
antiferromagnet on the linear chain provide a measure of the
effectiveness of the ECCM. However, we
know in advance that the nature of the transition
in the linear chain is very subtle, and quantum
 fluctuations present in this case are known to 
destroy N\'{e}el LRO completely at the 
isotropic point\cite{ba:bax}. 
Therefore, {\it a priori}, one would expect the ECCM to be 
more effective
for the square-lattice case, where the ground state will be closer
 to the classical uncorrelated state.

Numerical results for the ground states of models in one and two dimensions 
are expected to be qualitatively 
different \cite{mer:wag}.
By contrast with the NCCM, which has previously been applied, for example,
 to the 
spin-half {\it XXZ} model \cite{bi:pa} and the non-linear sigma model
\cite{no:ni}, the ECCM is expected to yield numerical evidence 
of this difference. 
As similar calculations are performed for any lattice dimensionality 
in the CCM, the qualitatively different behaviour 
of the solutions that we report in this paper for the two cases using the ECCM
 is not simply an artefact of the 
technique.

In a diagrammatic implementation of both CCM techniques, the
 ECCM, at any level of truncation,
 produces diagrams in greater
 abundance and of greater complexity than the NCCM, as 
shown in Sec. \ref{da}.
 Unlike in its NCCM counterpart, 
 it is possible within the ECCM formalism to define a 
spin-spin correlation function 
which is fully consistent with the corresponding definition 
of the magnetic order parameter, at {\it all} orders of truncation. 
Such properties of the ECCM make it attractive for describing correlation 
effects in physical systems, and particularly to study their quantum 
phase transitions and quantum order.

The CCM has been used very successfully \cite{{coe:fc},{ci:ciz},{ku:luh},{rfb:luh},{bi:kum},{br:be},{bi:rbi},{bi:ss}} to calculate 
the zero-temperature properties of a wide variety of extended many-body systems, including, for example, atoms 
and molecules, nuclear matter 
and finite nuclei, the
 electron gas, lattice gauge field theory,
 as well as spin-lattice systems 
\cite{{bi:pa},{ro:he},{bi:ha},{ze:st},{far:ze},{kr:fa}}.
In this paper, we present evidence from numerical results that 
demonstrates the particular superiority of the ECCM over the NCCM, 
in practice, 
to study such global properties
of spin systems as their quantum (zero-temperature) phase transitions.

\section{Extended Coupled Cluster Method Formalism}
Since detailed descriptions of the CCM
 formulation have been given elsewhere
 \cite{{arp:rpo},{arp:paj},{rb1:ja},{rb2:ja},{rb3:ja},{coe:fc},{ci:ciz},{ku:luh},{rfb:luh},{bi:kum},{br:be},{bi:rbi},{bi:ss}}, only the essential 
components will be given here.
Hubbard\cite{hu:hub} was one of the first to emphasize
 the importance of an exponential parametrization of the exact 
ground-state ket wave function $|\psi \rangle$ of an interacting 
many-body system, 

\begin{equation}
|\psi \rangle=e^S|\phi\rangle\ \ ;\hspace{5mm}S=\sum_I\!{\rule{0mm}{4mm}}^{'} s_{I}C_{I}^{\dagger}\ \ ,
\label{ket}
\end{equation}
in terms of a model state $|\phi\rangle$ which is not orthogonal to the 
exact ground-state ket
 wave function $|\psi \rangle$.
The correlation
operator $S$ in Eq. (\ref{ket}) is decomposed solely in terms of a
complete set of mutually commuting, linked, multiconfigurational creation operators $C^{\dagger}_{I}$ defined in terms of a complete set of 
many-body configurations \{$I$\}. These are, in turn, defined by a set-index $I$, which, for the spin-lattice systems under consideration, describes the set of spins which are flipped with respect to those contained in the suitably chosen, normalised, model state $|\phi\rangle$. The prime on the sum in equation (\ref{ket}) excludes the null set, $I\rightarrow 0$, corresponding to the identity operator, $C_0^{\dagger} \equiv1\hspace{-1.25mm} 1$.
We note that $C_I|\phi\rangle = 0$, $\forall I \neq 0$, 
and hence the state $|\psi\rangle$ obeys the intermediate normalisation 
condition, $\langle \phi|\psi\rangle=1$. 

Arponen\cite{arp:rpo} has shown how two distinct CCM parametrizations
 of the 
exact ground bra state $\langle\tilde{\psi}|$ can be given, which 
yield the NCCM and the ECCM respectively. The NCCM parametrization
of the ground bra state is given as:
\begin{equation}
\langle\tilde{\psi}|=\langle\phi|\tilde{S}e^{-S}\ \ ;
\hspace{5mm}\tilde{S}=1+\sum_{I}\!{\rule{0mm}{4mm}}^{'} 
\tilde{s}_{I}C_{I}\ \ ,
\label{bra1}
\end{equation}
whereas the corresponding ECCM parametrization is given as:
 \begin{equation}
\langle \tilde{\psi}|= 
\langle \phi | e^{S^{''}} e^{-S}\ \ ;
\hspace{5mm} S^{''}=\sum_I \!{\rule{0mm}{4mm}}^{'} s_I^{''}C_I\ \ .
\label{braa}
\end{equation}
Both $\tilde{S}$ and $S^{''}$ are
 constructed wholly in terms of multiconfigurational destruction operators, defined with respect to the model state $|\phi\rangle$. 
These are simply the hermitian adjoints, $C_I$, of the 
corresponding creation operators, $C_I^{\dagger}$, in Eq. (\ref{ket}). 
Furthermore, both parametrizations satisfy the normalization condition, 
$\langle \tilde{\psi} |\psi\rangle=1$. Although 
$\langle \tilde{\psi}|=\langle \psi|/\langle \psi|\psi\rangle$ 
formally, this relation may not be preserved when truncations are made, 
as explained below, in either parametrization. 

 The energy functional $\bar{H}_E$ in the ECCM formulation is thus 
given by:
 \begin{equation}
\bar{H}_E\equiv
\langle \phi |e^{S^{''}}e^{-S}He^{S}|\phi\rangle\ \ .
\label{sch}
\end{equation}
The similarity transformed
Hamiltonian $e^{-S}He^S$ may be expressed 
as the usual nested commutator expansion,
\begin{equation}
e^{-S}He^S=H+\left[H,S\right] +\frac{1}{2!} \left [\left [H,S \right],S \right]+\cdots\label{nest}\ \ .
\end{equation} 
Since $S$ is composed wholly of mutually commuting 
creation operators, the similarity 
transform in Eq. (\ref{nest}) only 
retains terms in which all amplitudes $s_I$ are linked to the Hamiltonian. 
Furthermore, provided that the Hamiltonian 
is finite order in the single-body operators,
 the expansion always terminates at 
finite order. Therefore, once the 
correlation operator $S$ is 
approximated, no further truncations are necessary.

Formally, the bra-state parametrizations in both the ECCM and NCCM 
formulations still preserve hermiticity.
Explicitly, we have the relations,
\begin{equation}
\langle\phi|\tilde{S} = \langle\phi|e^{S^{''}}=
\frac{\langle\phi|e^{S^{\dagger}} e^S}{\langle\phi|e^{S^{\dagger}} e^S|\phi\rangle}\ \ ,
\end{equation}
which are a consequence of satisfying the normalisation condition 
$\langle\tilde{\psi} |\psi\rangle =1$.
Although the manifest hermiticity is usually 
sacrificed at any level of truncation, 
the distinct bra-state parametrizations 
produce fully linked expectation values in Eq. (\ref{sch}), 
as they incorporate the similarity transform, and both 
$S^{''}$ and $\tilde{S}$ are composed wholly of destruction 
operators.

The 
double exponential structure of the ECCM formalism implies that Eq.
 (\ref{sch}) can be expressed in terms of the amplitudes
$\{s_I,s^{''}_I\}$ via a double similarity transform, 
and that both of these sets of ECCM amplitudes
 are linked-cluster quantities.
However, the NCCM formalism only allows a single exponential structure 
in terms of the amplitudes $\{s_I\}$. Although the amplitudes 
$\{s_I\}$ are linked-cluster quantities, the amplitudes $\{\tilde{s}_I\}$ 
are not.

 An arbitrary expectation value $\bar{A}$ can thus be expressed in the 
ECCM and the NCCM respectively as:
\begin{eqnarray}
\bar{A}_E & \equiv & \langle\phi|e^{S^{''}}
e^{-S}Ae^S|\phi\rangle=\bar{A}_E
 \left( \{s_I\},\{s_I^{''}\}\right)\label{exp}\ \ ,\\
\bar{A}_N & \equiv & \langle\phi|\tilde{S}
e^{-S}Ae^S|\phi\rangle=\bar{A}_N
 \left( \{s_I\},\{\tilde{s}_I\}\right)\ \ .
\label{expect}
\end{eqnarray}

In practice, either version of the CCM can only be implemented by truncating the expansions
in Eqs. (\ref{ket})-(\ref{braa}) by retaining only 
a finite or infinite subset of the complete set of configurations indices 
$\{I\}$.
In this paper the SUB2-$n$ approximation scheme is employed, which 
retains all configurations with up to 
two-body correlations between spins which are no more than a certain
 distance apart, specified by the index $n$. 
The details of the approximation scheme are given explicity in 
Sec. \ref{da} where the diagrammatic approach is described.

All ground-state properties can be determined in the 
ECCM and the NCCM respectively, by the stationary principle:
\begin{equation}
\frac{\partial \bar{H}_E}{\partial s_I}=
0  =  \frac{\partial \bar{H}_E}{\partial s^{''}_I}\ \ ; \ \ 
\frac{\partial \bar{H}_N}{\partial s_I}=
0  =  \frac{\partial \bar{H}_N}{\partial \tilde{s}_I}\ \ ,
\label{stat1}
\end{equation}
in the case where $|\psi\rangle$ and $\langle\tilde{\psi}|$
 correspond to the ground state. We remark, however, that due to the lack of manifest hermiticity between $\langle\tilde{\psi}|$ and $|\psi\rangle$ 
at a given level of truncation, the resulting stationary values of $\bar{H}_E$ 
and $\bar{H}_N$ are not necessarily upper bounds to the ground-state energy.

We note that the NCCM equations at a given order can formally be extracted
 from the corresponding ECCM equations at the same order by truncating the 
expansion of the exponentiated correlation operator
 $e^{S^{''}}$ at first order and 
performing the substitution 
$ s_I^{''}\rightarrow \tilde{s}_I$. 
In practice, the resulting ECCM equations
 from a particular truncation scheme 
are highly non-linear and, therefore, of greater complexity
 than their NCCM counterparts. However,
 the diagrammatic representation of the
 formalism which we present here makes tractable the numerical implementation
of high-order approximation schemes.

Very importantly, observables in the ECCM, 
which quantify the global behaviour
of a system in terms of its long-range order, 
obey the cluster property. Thus, for example, we have the 
very general relation,
\begin{equation}
\lim_{{\bf | r - r^{'} |}\rightarrow\infty}
\overline{A_{{\bf r}}  B_{\bf r^{'}}} =
{\bar{A}_{\bf r}}{\bar{B}_{\bf r^{'}}}\ \ ,
\label{global}
\end{equation}
where $A_{\bf r}$ and $B_{\bf r^{'}}$ are single-body 
operators acting at defined sites ${\bf r}$ and ${\bf r^{'}}$, 
respectively.
This condition is preserved by the ECCM due to the exact multiplicative 
separability of both the bra-state and ket-state parametrizations
in the corresponding large-distance limit, however the index set $\{ I\}$
 is truncated.
Consequently, the long-range order of physical systems can be 
examined unambiguously via the ECCM parametrization of their 
correlation functions.   
As the NCCM bra-state parametrization is not multiplicatively separable, 
however, the physical condition in Eq. (\ref{global}) 
will not in general hold at arbitrary levels of truncation within this scheme.

\section{TRANSFORMATION OF THE SPIN-HALF {\it XXZ} MODEL HAMILTONIAN FOR THE LINEAR CHAIN AND SQUARE LATTICE}
\subsection{The CCM approach to the spin-half {\it XXZ} model Hamiltonian}
\label{sect3}The spin-half {\it XXZ} model Hamiltonian is given by:

\begin{equation}
H=+\frac{1}{4}\sum_{i,\rho}
\left[\Delta\sigma_i^z\sigma_{i+\rho}^z
+\sigma_i^x\sigma_{i+\rho}^x+
\sigma_i^y\sigma_{i+\rho}^y\right]\ \ ,
\label{ham}
\end{equation}
where $\sigma_i^{\alpha}$, $\alpha = x,y,z$, are the Pauli spin matrices, 
$\Delta$ is the anisotropy parameter, and the summation 
is over all the $N$ lattice sites denoted by 
$i$ and over each of the nearest-neighbour vectors denoted by $\rho$.
We note that the linear
 chain and the 
square lattice are both bipartite lattices which can be 
split into two identical sublattices, 
which we denote as the $A$- and $B$-sublattices. Thus, each nearest-neighbour site to a site on the $A$-sublattice is on the $B$-sublattice, and vice versa.

In the Ising limit, $\Delta \rightarrow \infty$,
the $\Delta$-dependent term 
in the Hamiltonian of Eq. (\ref{ham})
becomes dominant and the classical $z$-aligned N\'{e}el state 
is the 
eigenstate
that yields the lowest energy. The $z$-aligned N\'{e}el 
state has
nearest-neighbour spins ordered antiparallel to one another 
in the $z$-direction, as 
illustrated below:

\begin{equation}
|\phi\rangle  = \ \ \bigotimes_{k\in A} |\uparrow\rangle_{ k}
\ \ \bigotimes_{ l\in B}|\downarrow
\rangle_l\ \ ,
\label{Neel}
\end{equation}
in a notation in which the $z$-axis points vertically upwards. 
However, for all finite values of $\Delta $ 
the terms in the $x$- and $y$-directions 
in the Hamiltonian in Eq. (\ref{ham}) come into play, and the
 $z$-aligned N\'{e}el state is no longer an eigenstate
of the Hamiltonian. The ground state for even $N$ now consists of a 
particular linear 
combination of all possible configurations
with $N/2$ up-pointing spins
and $N/2$ down-pointing spins.
All the configurations apart from the classical state 
which are present in the exact state are considered to be quantum fluctuations upon that state.
 At the isotropic point the Hamiltonian in Eq. (\ref{ham}) becomes 
rotationally invariant, 
such that the expectation value of an arbitrary spin is the same for 
any direction.

For $-1<\Delta < 1$ the 
true classical ground state can be any one of an
infinite number of degenerate N\'{e}el states with nearest-neighbour spins 
restricted to align antiparallel to one another in any direction 
in the {\it XY} plane. Schematically, we write:
\begin{equation}
|\phi\rangle  =\ \  \bigotimes_{k\in A}|
\leftarrow\rangle_k\ \ \bigotimes_{l\in B}|\rightarrow\rangle_l\ \ ,
\label{planar}
\end{equation}
\noindent choosing, say, the $x$-direction to be the alignment axis, 
and in a notation in which the positive $x$-axis points horizontally 
to the right. 
We note that there
 exists a trivial transformation of the Hamiltonian in Eq. (\ref{ham}) 
at the phase transition point, $\Delta=-1$, between the {\it XY}-Heisenberg 
and ferromagnetic phases (henceforth referred to as the ferromagnetic point), 
with the $x$-aligned N\'{e}el state as eigenstate, 
to the ferromagnetic Heisenberg Hamiltonian,
 with the $x$-aligned ferromagnetic 
state as eigenstate. 
Therefore, at $\Delta=-1$ the $x$-aligned N\'{e}el state is the 
true ground state and this point is chosen to be the initial point of the 
CCM calculation in the entire regime $-1<\Delta< 1$.
One imagines that the 
$x$-aligned N\'{e}el model state is 
close to the true 
ground state in the neighbourhood of this point.

\subsection{Transformation of the Hamiltonian 
in terms of a canted model state}In order to produce a ground-state solution that can describe 
both the {\it XY}-like ($-1<\Delta<1$) and 
Ising-like ($\Delta >1$) regions of the {\it XXZ} model, 
a canted model state is introduced. 
The canted model state, $|\phi_c\rangle\equiv|\phi(\xi_A,\xi_B)\rangle$,
 consists of spins on the $A$-sublattice
 pointing in one particular direction and spins on the $B$-sublattice
 also pointing in another particular direction, defined by the spin-half 
spin vectors $\xi_A$ and $\xi_B$, respectively.

In order to calculate the expectation 
values of arbitrary observables in the ECCM, 
an arbitrary rotation of the local spin axes 
is performed about the $y$-axis 
on each sublattice resulting in a notional 
rotation of the spins in the canted model state to the 
down position in the direction of the negative 
$z$-axis, 

\begin{equation}
U|\phi(\xi_A,\xi_B)\rangle  = \bigotimes^N_{ i=1 } |\downarrow\rangle_{ i}
\equiv|F\rangle \ \ ,
\label{down}
\end{equation}
where $U$ is a product of unitary matrices, which causes 
the spins in the canted model $|\phi_c\rangle$ to undergo 
a passive rotation such that they all point downwards in the rotated local frames.
Hence, the ECCM expectation value for an observable $A$ in Eq. (\ref{exp})
can be expressed with respect to the 
unrotated canted model state in the form,

\begin{equation}
\bar{A}=\langle\phi_c|U^{\dagger}e^{S^{''}}
e^{-S}\left(U A U^{\dagger}\right)e^S U|\phi_c\rangle\ \ .
\label{expect2}
\end{equation}
The rotation matrices for the canted model 
state on the $A$-sublattice and $B$-sublattice
 are given by:
\begin{equation}
U_{J}\equiv \exp\left(-i\theta_J\frac{\sigma^y}{2}\right)  = \cos\left(\frac{\theta_{J}}{2}
\right) \openone - i\sin\left(\frac{\theta_{J}}{2}\right)
 \sigma^y\ \ ;\ \  J=A,B \ \ ,
\label{ub}
\end{equation}
where each rotation is chosen to be performed about the $y$-axis in the 
{\it XZ}-plane. This involves no loss of 
generality, since all directions in the {\it XY}-plane are equivalent, due 
to the $U(1)_{xy}$ symmetry of the Hamiltonian in Eq. ({\ref{ham}).

Now, the transformed Hamiltonian arising from Eq. (\ref{ham}) can be 
written as:

\begin{eqnarray}
H^T \equiv  U H U^{\dagger} & = & +\frac{1}{2}\sum_{\langle k,l \rangle}
\left[\Delta \left(U_A\sigma_k^z U_A^{\dagger} \right) \left(U_B\sigma_{l}^z U_B^{\dagger} \right)
+\left(U_A\sigma_k^x U_A^{\dagger} \right) \left(U_B
\sigma_{l}^x U_B^{\dagger} \right)\right.\nonumber\\
     &  &  \left.+ \left(U_A\sigma_k^y U_A^{\dagger} \right) \left( U_B
\sigma_{l}^y U_B^{\dagger} \right) \right]\ \ ,\ \ k \in A,\ l\in B\ \ , 
\label{hamt}
\end{eqnarray}
where $k$ and $l$ are nearest neighbour vectors.

There are two degrees of freedom, $ \theta_A$ and $\theta_B$,
 present in the transformed Hamiltonian $H^T$, which are conveniently 
expressed via an equivalent set of relative and total orientation 
parameters,
\begin{equation}
\alpha\equiv\left(\theta_B - \theta_A \right)\ \ ; \hspace{5mm}
\beta\equiv - \left(\theta_A + \theta_B \right)\ \ .
\end{equation}
Particular values of $\alpha$ and $\beta$ lead to specific canted model states, as shown in Fig. 1. 
The single-spin 
creation and destruction operators are defined as $\sigma^{\pm}\equiv
\frac{1}{2}(\sigma^x\pm i\sigma^y)$ for all sites 
on the lattice, once the spins
 have all been rotated into the down position, 
as given in Eq. (\ref{down}).
Thus, $H^T$ in Eq. (\ref{hamt})
can be expressed in the form:
\begin{equation}
H^T   =  +\frac{1}{8} \sum_{i,\rho}\sum_{p,q}
T_{pq} \sigma_i^{p} \sigma_{i+ \rho}^{q}\ \ ,
\label{tpm}
\end{equation} 
where $p,q \in \{ z,+,- \}$ and $T_{pq}$ are functions of $\alpha,\beta$ and 
$\Delta$. Due to the hermiticity of the Hamiltonian the
factors $T_{pq}$ in equation (\ref{tpm})
 satisfy the relation, $T_{+z}=T^*_{-z}$.
 Moreover, the 
spatial symmetry
with respect to which lattice sites the spin operators
act upon, yields the relations:
\begin{equation}
 T_{z+} = T_{+z} \ \ ;\hspace{5mm}
 T_{z-} = T_{-z}\ \ ;\hspace{5mm} 
 T_{+-} = T_{-+}\ \ .
\end{equation}
Finally, the transformed Hamiltonian with 
a canted model state can be expressed as:

\begin{eqnarray}
H^T   =  +\frac{1}{8} \sum_{i,\rho}&&\left\{
 \frac{1}{2}\left[\left(1 + \Delta \right)\cos\alpha + \left(
1- \Delta \right)\cos\beta-2\right] \left( \sigma_i^+ \sigma_{i+ 
\rho}^+ + \sigma_i^- \sigma_{i+ \rho}^- \right)\right.
\nonumber\\
& &
+ \frac{1}{2}\left[\left(1+\Delta\right)\cos\alpha + \left(\Delta -1\right)\cos\beta\right] \sigma_i^z \sigma_{i+ 
\rho}^z 
\nonumber\\ 
& & + \left[\left(1+\Delta \right)\cos\alpha + \left(1  - \Delta \right)
\cos\beta+2\right]\sigma_i^+ \sigma_{i+ \rho}^-
\nonumber\\ 
& & 
+\left. \left(1-\Delta \right)\sin\beta \left(\sigma_i^z 
\sigma_{i+ \rho}^+ +
\sigma_i^z \sigma_{i+ \rho}^- \right)\right\}\ \ .
\label{tp}
\end{eqnarray}

\subsection{A mean-field calculation}
A mean-field calculation provides a guideline 
for the model-state analysis in the correlated CCM calculation, which will 
be performed in Sec. \ref{numres}.
We restrict our mean-field calculation here to finding an optimal 
state of the form of the canted model state 
$|\phi\left(\xi_A,\xi_B\right)\rangle$, in which no other correlations 
are present apart from those between the two sublattices implied by the angles 
$\alpha$ and $\beta$.
Consequently, the mean-field energy $E$, which is given by:
\begin{equation}
E = \langle \phi_c | H | \phi_c \rangle=\langle F | H^T | F\rangle\ \ ,
\label{En}
\end{equation}
yields the following expression:
\begin{equation}
\frac{E}{N} = \frac{z}{16}\left[\left(\Delta +  
1 \right)
\cos\alpha + \left(\Delta - 
1 \right)
\cos\beta\right]\ \ ,
\end{equation}
where $N$ is the number of lattice sites and $z$ is the co-ordination 
number of the lattice. 
The factor $T_{zz}$ in the Hamiltonian in Eq. (\ref{tpm}) 
yields the only non-zero, diagonal 
contribution from Eq. (\ref{En}).
The stationary values of $T_{zz}$ 
with respect to the angles $\alpha$ and $\beta$ 
yield the solutions for the energy $E$, which are shown in Fig. \ref{mean}.
Figure \ref{mean} also shows that the model states 
which make up the ground state in the mean-field case 
are the $z$-aligned N\'{e}el state for $\Delta\geq 1$, 
the $x$-aligned (or $y$-aligned) N\'{e}el state 
for $-1\leq\Delta\leq 1$, and the $z$-aligned ferromagnetic state for 
$\Delta < -1$.

\section{The Diagrammatic Approach}
\label{da}
\subsection{The Diagrammatic Technique}
The terms arising from a CCM calculation have
 been represented diagrammatically in previous work 
\cite{{ro:he},{ha:har}}.
For example, 
Roger and Hetherington\cite{ro:he} represent the terms arising from the NCCM 
energy functional diagrammatically simply for convenience. 
Unlike our representation, shown in Fig. \ref{rep},
 their diagrams do not include the
 destruction operators from the bra-state parametrization. Hence, diagrams are retained in Ref. \cite{ro:he}, which do not contribute to the energy functional, as shown by the NCCM diagrams in Fig. \ref{genmod1}. 
On the other hand, 
Harris\cite{ha:har} uses a representation where the 
algebra for 
performing the CCM with respect to spin states 
is reformulated in terms of particle-hole states. 
This particular representation was employed for the purpose of 
identifying cancelling terms and to aid the simplification and 
systematization of the algebra.

By contrast with these earlier diagrammatic approaches, the
 originality of the diagrammatic approach \cite{no:ni} used 
in the present work 
appears at the level of the numerical calculations. Since,   
the diagrammatic representation requires less computer memory and 
CPU time than the algebraic representation, 
higher-order truncation schemes can be implemented. 
There are several other advantages of using this particular
 approach: applying the formalism is straightforward;
 the terms which contribute the most in any 
calculation can readily be pinpointed; and the correlation function and 
sublattice magnetization can be expressed as
 sets of the diagrams 
which appear in the CCM energy functional arising
from $H^T$. 

Our diagrammatic technique involves casting the 
combinations of creation and destruction operators which appear 
in the CCM in 
a diagrammatic form, as shown in Fig. \ref{rep}.
The CCM SUB$n$ approximation scheme retains all configurations up 
to and including $n$-body correlations. Therefore, the SUB2 scheme
retains all the one-body correlation terms, denoted 
by $S_1$ in Fig. 3, that are allowed to arise from the 
terms in the Hamiltonian $H^T$ in Eq. (\ref{tp})
 which contain a single-body creation or destruction operator, 
as well as all 
the two-body correlations, denoted by $S_2$ in Fig. \ref{rep}.
 The SUB1 scheme retains only the one-body correlations. 
In the present work, the correlation operators $S$ and $S^{''}$ 
defined in Eqs. 
(\ref{ket}) and (\ref{braa}) are truncated via the SUB2 scheme as shown in Fig. \ref{rep}. 

As the equations that arise
 from the SUB2 scheme are not analytically soluble, a further restriction 
is placed on the maximum range of the two-body correlations from the 
SUB2 scheme. Therefore, the partial 
SUB2 approximation known as the SUB2-$n$ scheme is implemented. In the 
current diagrammatic formulation the SUB2-$n$ scheme
 retains only those 
two-body 
correlation coefficients $b_r$, where $r$ runs over the $n$
distinct vectors within a reference box of given size.
The vectors are defined to be distinct after allowing for all 
lattice symmetries. 
For example, the linear-chain case has the property that all correlation 
coefficients arising from vectors of the same length are identical.
 However, for the square-lattice case,
 vectors of the same 
length which are not otherwise identical under the lattice symmetries
yield distinct correlation coefficients. 
For example the vectors $(5,0)$ and $(3,4)$ have the same 
length but are distinct.

The one-body terms in $S_1$ from the SUB1 scheme
 simply transform a particular model state $\phi$ to an improved 
model state $|\phi^{'}\rangle$, where $|\phi\rangle$
 is in the neighbourhood of 
$|\phi^{'}\rangle$, via the transformation $|\phi^{'}\rangle=e^{S_1}|\phi
\rangle$.
The one-body terms from the SUB1 scheme 
are rigorously zero when the chosen 
model state is the same as the ground-state mean-field states, as the 
SUB1 scheme corresponds to a mean-field calculation.
Thus, the particular cases of the $z$-aligned and $x$-aligned 
N\'{e}el model states for $\Delta\geq 1$ and $-1\leq\Delta\leq 1$ respectively, as shown in 
Fig. \ref{mean}, from the Hamiltonian $H^T$ in the local 
rotated axes in Eq. (\ref{tp}),
yield rigorously zero one-body coefficients.
More general canted model states 
yield very small one-body contributions. 
Therefore, in order to keep the number of diagrams under control, we restrict the one-body terms to give, at most, a linear contribution, which 
corresponds to the NCCM SUB1 scheme. Hence, the restricted 
SUB2 bra-state parametrization 
can now be expressed as:

\begin{eqnarray}
\langle\tilde{\psi}| & = & \langle\phi|\left(1+S_1^{''}\right)\exp\left[S_2^{''}\right]\\
    & = & \langle\phi|\left(1+\sum_i k^{''}\sigma_i^- \right)\exp\left[\sum_{i,r} 
b_r^{''} \sigma_i^- \sigma_{i+r}^- \right]\ \ .
\end{eqnarray}

Once the energy functional (\ref{sch}) arising from 
$\bar{H}^T$ is calculated, the ground-state properties
 can be derived by applying the 
stationary principle of Eq. (\ref{stat1}).
An example of how the diagrammatic calculation is performed 
can be shown by taking the expectation value of the term 
$\sigma_i^z \sigma_{i+\rho}^z$, which is a constituent of $H^T$. The
nested commutator in Eq. (\ref{nest}) yields the following
 expression:
\begin{equation}
e^{-S}\sigma_i^z \sigma_{i+\rho}^z e^S =
\sigma_i^z \sigma_{i+\rho}^z + \left[\sigma_i^z \sigma_{i+\rho}^z,S\right] + \frac{1}{2}\left[\left[\sigma_i^z \sigma_{i+\rho}^z,S\right],S\right]\ \ .
\label{nestcom}
\end{equation}

The correlation operator, $S_2$ in Fig. \ref{rep},
which is described as a ket line, flips two spins in the
 model state $|F\rangle$ to the up position.
The nested commutator in Eq. (\ref{nestcom}) only
allows non-zero contributions, if the ket line
is connected to the term $\sigma_i^z \sigma_{i+\rho}^z$,
 in the Hamiltonian. Therefore, at
least one of the spins is flipped on 
either of the lattice sites $i$ and $i+\rho$.
Consequently, a spin is flipped on one other
site of the lattice.

The bra lines arising from $e^{S^{''}}$
 only yield non-zero contributions with the 
state described above if the final state 
is one
of all down-pointing spins. As the bra lines
flip up-spins to down-pointing spins, the
ends of the bra lines must be connected to ket lines or to a term in $H^T$ 
that contains creation operators. Otherwise, the 
destruction operators at either end of the bra line annihilate 
the state.
The combinatorics of both the bra lines and the 
ket lines in devising the diagrams is taken care of
by the CCM formalism. The ket lines arising
from the nested commutator and the bra lines
arising from $e^{S^{''}}$ have counting factors
that remove the possibility of obtaining 
diagrams which are equivalent under lattice symmetries.

\subsection{The diagrams arising from the ECCM formulation for 
macroscopic quantities}
\subsubsection{The diagrams arising from the energy functional}
The diagrammatic technique applied to the ECCM energy functional 
arising from the transformed Hamiltonian $H^T$ 
yields the diagrams shown in Fig. \ref{genmod1}.
\subsubsection{The sublattice magnetization}
The sublattice magnetization, or the order parameter in the Ising-Heisenberg and 
{\it XY}-Heisenberg phases, respectively, is defined by,
\begin{eqnarray}
M^z & = &\frac{2}{N} \sum_{k\in A}
|\langle  U^z\sigma_k^z U^{z\dagger}\rangle|
=-\frac{2}{N}\sum_{k\in A}\langle  \sigma_k^z \rangle\ \ ;\hspace{5mm}\Delta\geq 1\ \ ,\nonumber\\
M^x & = & \frac{2}{N} \sum_{k\in A}
|\langle  U^x\sigma_k^x U^{x\dagger}\rangle|
=-\frac{2}{N}\sum_{k\in A}\langle \sigma_k^z \rangle\ \ ;\hspace{5mm} -1\leq\Delta\leq 1\ \ .
\label{zr}
\end{eqnarray}
The inclusion of the minus sign ensures that $M^z$ and $M^x$ 
are positive in our rotated N\'{e}el basis, where 
$U^z$ and $U^x$ are the unitary operators, which rotate 
the local spin axes of the $z$-aligned and $x$-aligned N\'{e}el model states 
respectively to give the state $|F\rangle$.
The sublattice magnetization for $\Delta\geq 1$ is solely that of 
the $z$-aligned spins $M^z$, and for $-1\leq\Delta\leq 1$ the sublattice 
magnetization is that of the $x$-aligned 
spins $M^x$ in the present model state formulation.
The formal reason for this behaviour of the sublattice magnetization 
is given in Sec. \ref{syms}.

As the forms of the sublattice magnetizations $M^z$ and $M^x$ 
for the respective regions shown in Eq (\ref{zr}) are equivalent, 
they arise from the same diagrams, which are shown in Fig. \ref{diagmag}. 
Local quantities in the CCM formulation yield size-extensive results.
The diagrams for
 the ECCM sublattice magnetization are the same as those for the NCCM. 
Even though they are local quantities, the order parameters arise from 
summing correlations over the whole lattice,
 and in this way they are able to describe the ordering of individual spins 
throughout the system.

\subsubsection{The correlation function} 
In order to study critical phenomena it is 
important to study a particular correlation function which is 
consistent with the definition of the order parameter.
The 
spin-spin correlation function $g_0^{\alpha,\beta}$
 is defined by:
\begin{equation}
g_0^{\alpha,\beta}( r) \equiv \langle \sigma^{\alpha}_k 
\sigma^{\beta}_{k+r} \rangle\ \ ,\hspace{5mm} \alpha,\beta=x,y,z\ \ ,
\end{equation}
in terms of which we define the correlation function $g^{\alpha,\beta}( r)$ 
as follows,

\begin{equation}
g^{\alpha,\beta}( r) \equiv g_0^{\alpha,\beta}( r) - \langle \sigma^{\alpha}_k 
\rangle \langle \sigma^{\beta}_{k+r} \rangle\ \ .
\label{correlfn}
\end{equation}
where $r$ is restricted such that $k$ and $k+r$ are both on the $A$-sublattice. The translational invariance of the 
lattice implies, if $\alpha =\beta$ on a particular sublattice, 
$\langle \sigma^{\alpha}_i 
\rangle \equiv \langle \sigma^{\beta}_{i+r} \rangle$. Hence, 
the correlation function satisfies the fundamental property:
\begin{equation}
\lim_{r \rightarrow \infty}
g_0^{\alpha,\alpha}( r) = \left(\frac{2}{N} \sum_{k\in A}\langle \sigma^{\alpha}_k\rangle\right)^2
\equiv \left(M^{\alpha}\right)^2\ \ ,
\label{lim}
\end{equation}
where $M^{\alpha}$ is the order parameter and $i$ is summed over the sublattice of $N/2$ lattice sites. 
The 
expression for $g_0(r)$ in terms of 
the rotated canted model state is given by:

\begin{equation}
g_0(r)\equiv \langle U\sigma_k^x \sigma_{k+r}^x U^{\dagger} \rangle +
       \langle U\sigma_k^y \sigma_{k+r}^y U^{\dagger} \rangle +
       \langle U\sigma_k^z \sigma_{k+r}^z U^{\dagger}\rangle\ \ ,
\label{g0}
\end{equation}
where $U$ is the unitary operator in Eq. (\ref{ub}).
Clearly the form of the expression in Eq. (\ref{g0}) is similar to the form
of the corresponding energy 
functional. 
The same diagrams arise for
$g_0(r)$ as for the energy functional, which is shown in Fig. \ref{genmod1}.
 However, the links can now
be of arbitrary length, $r$. The property in Eq. (\ref{lim}),
implies that in the limit $r\rightarrow \infty$, surviving diagrams 
from $g_0(r)$
 can be interpreted as being topologically disconnected.
Conversely, Eq. (\ref{correlfn}) implies that the
correlation function $g(r)$ consists of the topologically connected diagrams 
arising from $g_0(r)$. Examining the topology of diagrams 
arising from the functional provides a straightforward way to deduce the
terms that yield the sublattice magnetization and the
 correlation function in an ECCM 
calculation.

The relevant correlation functions for the regions which encompass the 
Ising-Heisenberg and 
{\it XY}-Heisenberg phases respectively are given by:
\begin{eqnarray}
g^{zz}(r) & = & \langle U^{z} \sigma^z_k 
\sigma^z_{k+r} U^{z\dagger} \rangle - \langle U^z\sigma^z_k U^{z\dagger}
\rangle \langle U^{z}\sigma^z_{k+r} U^{z\dagger} \rangle\ \ ; \hspace{5mm}
\Delta\geq 1\ \ ,\nonumber\\
g^{xx}(r) & = & \langle U^{x}\sigma^x_k 
\sigma^x_{k+r} U^{x\dagger} \rangle - \langle U^x \sigma^x_k U^{x\dagger}
\rangle \langle U^{x} \sigma^x_{k+r} U^{x\dagger} \rangle\ \ ;\hspace{5mm}
-1\leq\Delta\leq 1\ \ .
\label{corspifn2}
\end{eqnarray}
In the limit $r \rightarrow \infty$, the 
correlations
between spins die off as shown in Eq. (\ref{lim}).
The spin-spin correlation function for the region $-1\leq\Delta\leq\infty$
has the form $\pm\langle\sigma_i^z \sigma_{i+r}^z \rangle$ as 
$\langle U^{z} \sigma_i^z \sigma_{i+r}^z U^{z\dagger} \rangle 
=
\langle U^{x} \sigma_i^x \sigma_{i+r}^x U^{x\dagger} \rangle= \pm\langle\sigma_i^z \sigma_{i+r}^z \rangle$, where the sign is positive if both lattice sites 
are on the same sublattice and negative otherwise. 
Therefore, the correlation function
consists of diagrams
similar to the connected diagrams arising
from the term $\langle\sigma_i^z \sigma_{i+\rho}^z\rangle $ 
 in 
the energy functional $\bar{H}^T$.
The diagrams arising from the correlation function in 
Eq. (\ref{corspifn2}) are shown in Fig. \ref{diagcor}.

The connected diagrams which appear in the NCCM correlation function
 are equivalent 
to those in Fig. \ref{diagcor} with the exception 
of those indicated in the box. 
However, there are also disconnected diagrams in the NCCM formulation
of the correlation function. This occurs because one of the diagrams 
that arises from the square of the 
sublattice magnetization in Fig. \ref{diagcor} is disconnected and corresponds to a 
higher-order term that cannot appear in the spin-spin correlation 
function, 
due to the NCCM bra-state parametrization. 
Consequently, the cancellation of the disconnected diagram cannot occur 
when the correlation function is calculated in Eq. (\ref{corspifn2}). 
 However, the ECCM  bra-state parametrization yields higher-order 
terms than the NCCM, which allows this cancellation to take place. 
Hence, the correlation function is solely composed of connected 
diagrams, which possesses the fundamental
 property of Eq. (\ref{lim}). In this way the importance of the
 cluster property, which is intrinsic in the ECCM formalism in 
Eq. (\ref{global}), can be seen for the correlation function.

\section{The symmetries of the mean-field model state}
\label{syms}
Criteria for a good 
choice of model state in terms of 
the maximum overlap between the model state and the ket state have 
been derived by 
K\"{u}mmel \cite{ku:kum}. One of these criteria specifies that there should be 
no contribution from the one-body terms (or the SUB1 scheme) 
in the CCM calculation. 
The model states arising from the ground-state mean-field calculation 
satisfy this criteria.
 From the 
canted model state, the mean-field calculation yields the ground state 
for the $z$-aligned N\'{e}el model state in the region $\Delta\geq 1$ and
the $x$-aligned N\'{e}el model state 
in the region $-1\leq\Delta\leq 1$, as shown in Fig. \ref{mean}. 
However, it can be shown, by applying a theorem first 
enunciated by Xian \cite{ya:xi} to 
these mean-field model states, that the ket state 
$|\psi\rangle$ cannot exhibit symmetry breaking at $\Delta=1$.

Xian's theorem \cite{ya:xi} states that the CCM equations provide at least one 
solution which guarantees the symmetry of the model state, if this 
symmetry is one of those belonging to the Hamiltonian. 
The theorem is proved by employing a symmetry operator $\Lambda$, 
which has the properties that the Hamiltonian in Eq. \ref{ham} 
commutes with $\Lambda$ and that the model state is an eigenstate 
of $\Lambda$.

In the case of the $z$-aligned N\'{e}el model state, the symmetry operator $\Lambda^z$ is given by the rotation operator,

\begin{equation}
\Lambda^z =\bigotimes^N_{ i=1 } \left[\cos\left(\frac{\chi}{2}\right)\openone + 
i\sin\left(\frac{\chi}{2}\right) \sigma^z_i\right]\ \ ,
\end{equation}
where $i$ runs over all lattice sites.
The order parameter for an arbitrary direction in the {\it XY}-plane
is defined by $M^{xy}\equiv M^x \cos\alpha +M^y\sin\alpha$. 
The action of the rotation operator  $\Lambda^{z}$ on the operators 
whose expectation values on the sublattice yield $M^{xy}$ 
is given by:
\begin{equation} 
\Lambda^{z \dagger}(\sigma^x \cos\alpha +\sigma^y\sin\alpha)\Lambda^z =
\sigma^x\cos\left(\alpha + \chi\right) +\sigma^y\sin\left(\alpha + \chi\right)\ \ .
\label{mxy}
\end{equation}
The only solution that guarantees the required invariance of 
$M^{xy}$ under rotation 
is $M^x\equiv 0$ and $M^y\equiv 0$, such that the $U(1)_{xy}$ symmetry 
of the $z$-aligned N\'{e}el model state is preserved. Consequently, the theorem \cite{ya:xi}
shows that $|\psi\rangle$ cannot exhibit broken $U(1)_{xy}$ symmetry.

Similarly, the discrete symmetry operator for the particular case of the 
$x$-aligned N\'{e}el model state is given by:
\begin{equation}
\Lambda^x=\bigotimes^N_{ i=1 }\ \sigma^x_i\ \ .
\end{equation} 
The transformation of the order parameter $M^z$ to $-M^z$ 
with $\Lambda^x$ implies 
that $M^z\equiv 0$. Thus, the $Z(2)$ symmetry of the 
$x$-aligned N\'{e}el model state is preserved and consequently 
$|\psi\rangle$ cannot exhibit broken $Z(2)$ symmetry.
Hence, it follows that the $z$-aligned 
N\'{e}el model state only describes the region $\Delta \geq 1$, where the 
ground state does not have broken $U(1)_{xy}$ symmetry.
Similarly, the $x$-aligned N\'{e}el model 
state only describes the region $-1\leq\Delta\leq 1$, 
where the ground state does not have broken $Z(2)$ symmetry.

\section{NUMERICAL RESULTS}
\label{numres}The diagrams in the ECCM energy functional 
corresponding to the Hamiltonian $H^T$, which arise from 
the SUB2-$n$ approximation scheme in Fig. \ref{genmod1}, are used to 
obtain the numerical results for both the linear-chain and 
square-lattice cases. 
The qualitative behaviour of the numerical 
results differs for the two cases. 
This difference can only arise from the 
different multiplicities of the diagrams, which themselves are due
 to the symmetries of the lines
 in the diagrams differing in these cases.

From the implementation of the NCCM SUB1 scheme it can be seen that 
the coefficients $k$ and $k^{''}$ provide a negligible contribution 
to the overall calculation, where they are not rigorously zero for 
the $x$-aligned and $z$-aligned N\'{e}el model states. Therefore, with 
the benefit of hindsight, implementing the ECCM SUB1, which only introduces 
additional higher-order terms, would not be much different to the NCCM SUB1.

\subsection{The spin-half {\it XXZ} model on the linear 
chain}There are both exact results for the case of the linear chain and more appropriate
 numerical techniques than the CCM, such as the density matrix renormalization 
group and spin-wave theory, for examining more general one-dimensional spin 
chains. However, the presence of exact results makes the linear
chain an important case to gain insight 
into the behaviour of the numerical results from the approximated ECCM 
formalism.

The mean-field calculation shows that both the 
$z$-aligned and $x$-aligned N\'{e}el model states introduce 
complete $z$-aligned and $x$-aligned N\'{e}el ordering respectively,
 at the isotropic point.
 Figures \ref{1dmag} and \ref{1dmagXY} show that the ECCM SUB2-$n$ scheme 
fails to pick up 
all the quantum fluctuations in the physical system that result in the LRO
 being completely destroyed at the isotropic point. However, the 
exact order 
parameter from the
 linear-chain case exhibits subtle behaviour, which, in principle,
 would require 
the implementation of extremely high-order ECCM approximation schemes.

The ECCM numerical solutions, for the quantities $M^x$ and $E_g$, 
from truncation schemes of orders higher than 
SUB2-2 with the
$x$-aligned N\'{e}el model state, at the initial point $\Delta=-1$, 
do not exist for $\Delta > -1$, as shown in Figs. \ref{1dmagXY} and \ref{highen}.
This occurs for any scheme higher than 
SUB2-2 because there is a degeneracy in the numerical 
solutions 
at $\Delta=-1$, which makes it a poor choice of starting point. 
An alternative choice of initial point for the CCM calculation is the 
isotropic point, since the rotated Hamiltonians obtained from both 
the $x$-aligned and $z$-aligned N\'{e}el model states
 are equivalent at this point.
These solutions turn back and become unphysical. 
The points at which the solutions
 become unphysical are known as the terminating points, and 
are shown in 
Figs. \ref{1dmagXY} and \ref{highen}.

The terminating points from the SUB2-$n$ scheme 
based on the $x$-aligned N\'{e}el model state, at the orders that
have so far been obtained, converge weakly to $1.025\pm 0.005$ as
$n\rightarrow \infty$, which can be interpreted as the SUB2 terminating point.
This result was obtained from the best possible
 mean square fit of $1/n^{0.563}$. It 
indicates that the $x$-aligned N\'{e}el
 model state from the SUB2 approximation scheme would not
 yield a solution in the {\it XY}-like region, where it is considered
 to be a better model state than the $z$-aligned N\'{e}el model-state. 
This is surprising since higher orders include more of the long-range 
correlations in the system, and consequently should reveal more 
of the physical behaviour of the system. The exact 
ground state of the 
linear chain in the {\it XY}-like region possesses $U(1)_{xy}$ symmetry, 
which indicates 
that the $x$-aligned N\'{e}el model state, 
as the sole input in the technique,
breaks this symmetry causing solutions to terminate in the region $\Delta<1$.
However, the numerical results from the ECCM 
for the ground-state energy and sublattice magnetization $M^x$ 
are good at low orders.

Figure \ref{limen} shows that the low-order ECCM SUB2-$n$ solutions 
have the same qualitative behaviour
as the exact ground-state energy solution, where a maximum deviation 
in the energy occurs 
at the isotropic point. Numerical results from the NCCM LSUB$n$ 
scheme \cite{far:ze} yields a
numerically accurate
ground-state energy at the isotropic point from an 
extrapolation, as shown in Table. \ref{1dtaben}. The LSUB$n$ 
scheme retains only those configurations in the correlation operator 
$S$ which contain any number of spin flips with respect to the 
model state over a localized region of $n$ contiguous sites, and 
which are compatible with the restriction $S_T^z\equiv\frac{1}{2}
\sum_{i=1}^N\sigma_i^z =0$. We note that this restriction follows 
from the facts that $S_T^z$ commutes with $H$ and that the ground 
state is expected to be in the $S_T^z=0$ sector.

\subsection{The spin-half {\it XXZ} model on the 
square lattice}There are no exact results for the square-lattice case, which 
therefore presents a significant 
challenge for numerical techniques, particularly for the study 
of the nature of 
the expected transition at the isotropic point. Unlike the linear-chain case, 
the square-lattice case is believed to possess LRO at the isotropic point.
 Evidence 
for this comes from a number of techniques as shown in Table \ref{hctechs}.
The quantum fluctuations in the square-lattice case should be 
smaller than those in the linear-chain case, due to the 
higher co-ordination number of the lattice. For this 
reason one expects that the CCM should be better at deducing the physical behaviour of 
the square-lattice case.

The numerical solutions for $M^z$ and $E_g$, 
from truncation schemes at orders higher than SUB2-12 
with the
$z$-aligned N\'{e}el model state, terminate before 
the isotropic point, as shown in 
Figs. \ref{2dneelmag} and \ref{2den} respectively.
 The SUB2 terminating point of the CCM calculations based on the $z$-aligned 
N\'{e}el model state 
can be accurately determined 
by an extrapolation, where the best mean square fit of 
$1/n^2$ gives a SUB2 terminating 
point at $\Delta\approx 1.03903\pm 0.00077$. 
Since the $z$-aligned N\'{e}el model state only breaks the discrete 
$Z(2)$ symmetry of the ground state, the SUB2-$n$ terminating points
exhibit this good convergence.

Similarly, solutions starting at $\Delta=-1$ from the SUB2-20 scheme
 with the $x$-aligned N\'{e}el 
model state for $E_g$ and $M^x$ terminate 
in the {\it XY}-like region, as shown in Figs. \ref{2den} and 
\ref{2dxymag} respectively. 
The SUB2-20 approximation scheme is 
the highest we have so far investigated, as well as being the only 
scheme to yield a terminating solution in the {\it XY}-like 
region. By contrast, at lower orders terminating points occur for $\Delta>1$. 
Therefore, although the SUB2 terminating point cannot be accurately determined,
it is expected to be closer to $\Delta=-1$ than the 
SUB2-20 terminating point at $0.78$. Consequently,
 it is not possible to obtain a solution from high-order SUB2-$n$ schemes 
that can describe the 
neighbourhood of the isotropic point on the $\Delta<1$ side.

Unlike in the linear-chain case the existence of a solution from high-order 
schemes with the $x$-aligned N\'{e}el model state in the square-lattice 
case indicates that the continuous $U(1)_{xy}$ symmetry of the 
ground state is broken.
 However, the bad convergence of the terminating points from the $x$-aligned 
N\'{e}el model state in the square-lattice case may indicate that the 
ground state is almost symmetric and, hence, a model state with 
$U(1)_{xy}$ symmetry would be a more appropriate 
starting point for a CCM calculation.

 The SUB2-12 scheme is the 
highest order SUB2-$n$ 
scheme that yields the sublattice magnetization $M^z$ at the isotropic point, 
because at higher orders the solutions terminate before the isotropic point.
The SUB2-12 scheme yields a value of $M^z\approx 0.689$ of the saturation value. 
This is an 
improvement upon the NCCM SUB2 value of $M^z\approx 0.81$, and the 
best NCCM result from the LSUB8 scheme \cite{far:ze}, which yields a value of
 $M^z\approx 0.705$, 
when compared to 
other numerical techniques, as shown in Table. \ref{hctechs}. 
The extrapolated numerical value of the 
magnetization from
LSUB$\infty$, $M^z \approx 0.622$, 
is the result of a quadratic 
fit in $1/n$ with a restricted set of three data points. The best linear fit 
in $1/n$ with the same data points yields $M^z\approx 0.646\pm 0.002$

Quantitatively, the CCM yields accurate values for the ground-state energy 
$E_g$ compared to other numerical techniques as shown in Table. \ref{gsenergy}.
The NCCM SUB2 scheme \cite{bi:pa} at the isotropic point yields a ground-state energy of 
$E_g\approx -0.651$, which 
the ECCM SUB2 scheme improves upon, 
via an extrapolation with values that converge well, with a numerical 
result of $E_g\approx -0.667$.
The calculation of the ground-state energy from the 
NCCM LSUB$n$ schemes \cite{far:ze} are extremely accurate when compared to the best Monte Carlo result \cite{kr:kru} where the limit $n\rightarrow\infty$ 
is equivalent to including the complete set of many-body configurations. 

\subsection{The canted model state in the 
correlated ECCM SUB2-$n$ calculation}
The effect of the canted model state, which has two 
degrees of freedom, in a CCM calculation is examined here.
In practice, the spins in the canted model state 
are restricted 
to point in one direction on a particular sublattice. 
Nevertheless, for the spin-half {\it XXZ} model, the
CCM with the canted model state provides a good analysis of the 
effect of the model state in these calculations.
In terms of the numerical calculations, performing the CCM with the 
canted model state is a new approach, since 
the earlier CCM calculations for spin systems 
have chosen pre-determined model states with no free parameters, and 
with the choice based on either prior knowledge or classical behaviour. 
The use of the 
canted model state allows the extra freedom to partially tailor the model state to the specific value of the anisotropy parameter.
This is achieved by the application of the 
stationary equation (\ref{stat1}) and the variation with respect to the 
angles $\alpha$ and $\beta$, which define the model state, where 
the minimum energy is automatically 
chosen for $\alpha$ and $\beta$ at each value of the anisotropy $\Delta$.
Therefore, we employ this particular form for the numerical CCM calculation, 
in the hope that it will chose model states which lie closest to the 
true ground state.

\subsubsection{The behaviour of the canted model state with varying angles}
For the calculations with varying anisotropy, the angles $\alpha$ and 
$\beta$ are free to vary.
The model states that arise from the ECCM SUB2-$n$ calculation in 
this case are the same as those from 
the mean-field calculation, as shown in Fig. \ref{mean},
 for $-1\leq\Delta\leq \infty$. This is also illustrated by 
Fig. \ref{modst} in the neighbourhood of the isotropic 
point. As discussed in Sec. \ref{syms}, 
these mean-field model states yield the CCM ground state 
that preserves either the $U(1)_{xy}$ or $Z(2)$ symmetry of the 
Hamiltonian.

K\"{u}mmel's work\cite{ku:kum} provides a guideline to examine the 
effect of the mean-field 
model states in the CCM calculation, if translational invariance of the rotated model state $|F\rangle$ is assumed. 
It can be shown that the overlap condition 
for the mean-field model states is maximized if the sum, over the whole lattice, of the 
two-body correlations, which is defined in Fig. \ref{rep} as $S_2$,
 obey $\sum_r b_r <1/2$. 
This result is extended to show that $\sum_r b_r \rightarrow 1/2$ 
indicates a phase transition point. As $\Delta\rightarrow 1$ 
this condition holds for the $z$-aligned N\'{e}el model state in the 
linear-chain case and only approximately holds for the square-lattice 
case. For the square-lattice case, this may indicate the numerical 
instability of the solution as $\Delta\rightarrow 1$ and consequently 
the need for a model state with greater overlap in this region.

The solution for the energy at the isotropic point 
has a continuous degeneracy in terms of the angle $\beta$, 
for a fixed value of $\alpha =\pi$ which 
yields the set of all model states with N\'{e}el symmetry.
 However, away from the isotropic point there 
is a unique minimum in the
 solution. Therefore, in practice, by choosing a fixed angle $\beta$, we break the 
$SU(2)$ symmetry at the isotropic point. In doing so, LRO is introduced 
in the direction specified by the angle $\beta$, which is diminished by the 
correlations from a CCM calculation.

\subsubsection{The behaviour of the canted model state with fixed angles}
The effect of particular model states that require 
the inclusion of one-body terms in the CCM calculation are 
examined here. 
In the Ising-like region, for the angle $\beta$ set to $\pi$, the angle 
$\alpha$ is fixed 
to various values
 which define model states in 
the neighbourhood of the $z$-aligned N\'{e}el state, 
i.e., $\alpha\sim\pi$. 
As $\alpha$ is decreased, 
each numerical solution for the energy in the square-lattice case, 
from the highest-order ECCM SUB2-$n$ scheme, 
is successively higher and terminates 
earlier than the previous solution. Furthermore, 
for model states defined by 
$\alpha \leq\pi/2$, a solution cannot be recovered.

These non-mean-field model states, in the neighbourhood of the 
$z$-aligned N\'{e}el model state, require one-body terms in the 
CCM calculation. Hence, the ground state can exhibit 
broken $U(1)_{xy}$ symmetry\cite{ya:xi}. However, these model states are 
bad choices in comparison to the symmetry-preserving
 mean-field model states, due to the poor qualitative and 
quantative behaviour of the 
numerical solutions for the energy. Moreover, due to the termination of these 
numerical solutions well before $\Delta=1$, broken $U(1)_{xy}$ symmetry 
does not physically occur in this region. Thus, the
 symmetry-breaking nature of these 
poor choices of model state cannot be utilised. 
It seems that non-mean-field model states with one-body correlations 
try to approach the mean-field model states and, in doing so, 
do a worse job at describing the system.

\subsubsection{The non-trivial 
behaviour of the canted model state in the
Ising limit}We now examine the effect of the canted model states in the Ising limit.
As shown in Fig. \ref{modst}, for the Ising-like region of the linear-chain 
case the lowest energy, and therefore the optimum
 model state to yield the ground-state energy, is the $z$-aligned N\'{e}el
 state. Initially, in the Ising-like region the periodicity of the solution 
for the ground-state energy in terms of the angle $\beta$ is $2\pi$, when
 $\alpha$ is fixed to $\pi$, as shown in Fig. \ref{mod} for the linear-chain 
case,
and with minima at $\pi$ mod($2\pi$). 
In the Ising limit the solution
asymptotically develops a periodicity of $\pi$ in the angle $\beta$. 
The states at $\beta=0$ mod($2\pi$),
 which define the $x$-aligned N\'{e}el state,
 become degenerate with those at $\beta=\pi$ mod($2\pi$), 
which define the $z$-aligned N\'{e}el state.
Interestingly, this analysis reveals that in the Ising limit,
the $z$-aligned N\'{e}el ground state is obtained from the 
the $x$-aligned N\'{e}el model state with large one-body 
coefficients $k$.

\section{Discussion and Conclusions}
The diagrammatic technique has made possible the implementation of 
high-order SUB2-$n$ schemes in the ECCM formulation, since it requires less
 computer memory and CPU time than performing the CCM with algebraic methods. 
This is because the multiplicties of the diagrams are only determined once 
and can thereafter be used throughout the calculation. 
For the case of the square lattice, accurate numerical results in 
comparison to other numerical techniques have been obtained for the 
region $-1<\Delta < \infty$. In particular, 
at the isotropic point the most accurate value for the order parameter of 
$M^z \approx 0.689$ for any non-extrapolated CCM calculation has been obtained. This 
suggests that the ECCM SUB2-$n$ scheme describes the wave function
 better than previous NCCM formulations at similar levels of truncation.

The qualitative behaviour of the numerical results indicate that 
the present CCM treatment requires further improvement for the study of phase transitions.
 In the case of the linear chain the ECCM fails to describe the 
system for $| \Delta | \leq 1 $. For the square-lattice case there is 
a failure to describe the region near to the isotropic point. Alternatively, the manner in which the numerical solutions become unphysical provides 
further knowledge of the system and the ECCM.

It is intuitively clear that a good choice 
of model state in a CCM calculation should be as close as possible to 
to the exact ground state or, alternatively, share the same underlying symmetries as 
the exact ground state, depending on the system under consideration.
The high-order SUB2-$n$ ECCM numerical results show that the important 
facet of a good choice of model state is dependent on the extent to which the system is ordered at the 
isotropic point and in the {\it XY}-like region.
The linear-chain case with both the $z$- and $x$-aligned N\'{e}el model states,
 and the square-lattice case with the $x$-aligned N\'{e}el model state, 
indicate that when either LRO does not exist or the value of the order parameter may be 
quite small, it is important to chose a model state which does not 
artificially break the symmetry of the Hamiltonian. 

Alternatively, the square-lattice case with the $z$-aligned 
N\'{e}el model state suggests that if the system has broken 
symmetry, the large quantum fluctuations that result from the 
massless excitations could also cause solutions to terminate.
In this case, it is more important to chose a model state 
which is as close as possible to the exact ground state. 
This problem of correlations becoming too large seems to occur 
generally in quantum many-body techniques.
Advantageously, introducing broken symmetry into such a system yields the 
benefit of not having to perform high-order calculations to approximate 
the true ground state well.

As shown above, our analysis, as it stands, requires prior knowledge of the 
system. 
Although other numerical techniques \cite{{an:and},{kan:yo},{og:og},{kr:kru},{si:sin},{zh:wh}} suggest that the square-lattice 
Heisenberg model
 possesses LRO, the value of the order parameter might be small. 
Therefore, the best CCM approach for this system is to 
define a model state with a flexible parametrization, 
so that it does not artificially break the symmetry 
of the Hamiltonian, and in so doing can give indications of possible 
symmetry breaking by examining the effect of correlations on the 
order parameter.
In order to achieve this end the model-state parametrization 
should yield a mean-field solution in which the order parameter 
smoothly and continuously tends to zero at the isotropic point. 
In this way the ECCM formulation can be extended 
over the whole regime to include the long-range correlations at the 
critical point.

\section*{Acknowledgments}
We thank S. Fantoni for a useful discussion.
One of us (RFB) gratefully acknowledges support for this work in the form 
of a research grant from the Engineering and Physical Sciences Research Council (EPSRC) of Great Britain.

\begin{table}

\caption{The ground-state energy per spin for the one-dimensional chain 
at the isotropic point, $\Delta=1$, 
under various CCM schemes, compared with the exact results.}

\begin{center}
 \begin{tabular}{|c|c|c|c|}\hline
  NCCM SUB2 & ECCM SUB2 & NCCM LSUB$\infty$  &  Exact \\ \hline
  $-0.419$ & $-0.433$ & $-0.443149$ & $-0.443147$ \\ \hline
 \end{tabular}
 \end{center}
\label{1dtaben}
 \end{table}
\begin{table}
\caption{The sublattice magnetization $M^z$ 
for the two-dimensional square lattice at the isotropic point, 
$\Delta =1$, 
as a fraction of the saturation value,
 from various CCM approximation schemes, a series expansion, 
a Monte Carlo simulation, and spin-wave theory (SWT). 
 }

\begin{center}
 \begin{tabular}{|c|c|c|c|c|c|}\hline
  ECCM SUB2-12 & NCCM LSUB8\tablenote{Reference \cite{far:ze}.} & NCCM LSUB$\infty^a$. & 
series expansion\tablenote{Reference \cite{zh:wh}.}  &  
Monte Carlo\tablenote{Reference \cite{kr:kru}.} & 
SWT\tablenote{Reference \cite{an:and}.} \\ \hline
  $0.689$ & $0.705$ & $0.622$ & $0.614\pm 0.002$ & $0.615\pm 0.005$ & 0.606\\ \hline
 \end{tabular}
 \end{center}
\label{hctechs}
 \end{table}
\begin{table}
\caption{The ground-state energy per spin for the two-dimensional 
square lattice at the isotropic point, $\Delta =1$, 
from various CCM schemes, compared with the results of a 
Monte Carlo simulation. 
	}
 \begin{center}
 \begin{tabular}{|c|c|c|}\hline
 ECCM SUB2 & NCCM LSUB$\infty\tablenote{Reference \cite{far:ze}.}$& Monte Carlo\tablenote{Reference \cite{kr:kru}.} \\ \hline
  $-0.667$ &$-0.66968\pm 0.00004$ & $-0.66934\pm 0.00003$\\ \hline
 \end{tabular}
 \end{center}
\label{gsenergy}
 \end{table}
 \begin{figure}
    \caption{Various model states canted in the {\it XZ}-plane, 
defined by the angles $\alpha$ 
and $\beta$, are shown in the original global co-ordinate frame, in 
a notation in which the positive $x$-axis points to the right and 
the positive $z$-axis points upwards. 
Note that the states with $\alpha =\pi$ are all N\'{e}el 
states and those with $\alpha =0$ are all ferromagnetic states.}
\label{ang2}
\end{figure}
\begin{figure}
   \caption{The stationary energy eigenvalues, $E/N$, as a 
function of the anisotropy, $\Delta$, for the 
mean-field case. The ground-state energy is denoted 
by a thick line.}
\label{mean}
\end{figure}  

\begin{figure}
   \caption{The diagrammatic representation of the ECCM formalism. 
The dashed lines denote the interaction terms from the Hamiltonian. 
The circle and the the cross denote respectively a creation operator and a 
destruction operator. The 
straight solid lines are known as the ket lines and the wiggly solid lines are 
known as the bra lines. }
\label{rep}
\end{figure}
\begin{figure}
   \caption{The diagrams arising from the 
ECCM energy functional (2.4) using the Hamiltonian 
$H^T$, which yields the ground-state energy. The factors denoted by the 
letters are composed of functions of the 
angles $\alpha$ and $\beta$ and the anisotropy $\Delta$. The factor $z$ 
denotes the co-ordination number of the lattice.
 The diagrams that appear up to the 
solid line are those from the NCCM formulation.}
\label{genmod1}
\end{figure} 
\begin{figure}
   \caption{The diagrams constituting the sublattice magnetizations $M^z$
and $M^x$. }
\label{diagmag}
\end{figure}
\begin{figure}
   \caption{The diagrams for the correlation functions, $g^{zz}$
and $g^{xx}$. }
\label{diagcor}
\end{figure}

\begin{figure}
   \caption{The sublattice magnetization in the $z$ direction, $M^z$, 
as a function 
of the anisotropy $\Delta$ for the linear-chain case, from the ECCM SUB2-$n$ schemes using the $z$-aligned N\'{e}el model state, a mean-field calculation, and the exact results.}
\label{1dmag}
\end{figure} 
\begin{figure}
   \caption{The sublattice magnetization in the $x$-direction, $M^x$, 
as a function 
of the anisotropy $\Delta$ for the linear-chain case, from the ECCM SUB2-$n$ schemes and a mean-field calculation with the $x$-aligned N\'{e}el model state.
 For truncation schemes higher than the SUB2-2, solutions 
turn back, so only the physical solution is shown with a solid circle indicating the turning back point, also known as the terminating point. 
It is known that there is no LRO over the whole region, $-1<\Delta <1$.}
\label{1dmagXY}
\end{figure} 
\begin{figure}
   \caption{The ground-state energy per spin, $E_g/N$, as a function 
of the anisotropy $\Delta$ for the linear-chain case from high-order 
ECCM SUB2-$n$ schemes using the $x$-aligned N\'{e}el model state, compared with
 the
exact results.}
\label{highen}
\end{figure}
\begin{figure}
   \caption{The ground-state energy per spin, $E_g/N$, as a function 
of the anisotropy $\Delta$ for the linear-chain case, from the low-order 
ECCM SUB2-2 scheme using the $z$-aligned and $x$-aligned N\'{e}el model states in the 
regions $\Delta\geq 1$ and $-1\leq\Delta\leq 1$ respectively, and the
exact results}
\label{limen}
\end{figure}  
\begin{figure}
    \caption{The sublattice magnetization in the $z$-direction, $M^z$, 
as a function 
of the anisotropy $\Delta$ for the square-lattice case, from the high-order 
ECCM SUB2-$n$ schemes using the $z$-aligned N\'{e}el model state. The results converge at high orders of the SUB2-$n$ 
truncation scheme, thus giving the full SUB2 result, which corresponds to $n\rightarrow\infty$.}
\label{2dneelmag}
\end{figure}

\begin{figure}
    \caption{The ground-state energy per spin, $E_g/N$, as a function 
of the anisotropy $\Delta$, from the $z$-aligned and $x$-aligned 
N\'{e}el model states in the regions $\Delta\geq 1$ and  
$-1\leq\Delta\leq 1$ respectively, for the square-lattice case. 
The terminating points of the 
solutions from the ECCM full SUB2 and SUB2-20 approximation schemes, 
in the regions $\Delta\geq 1$ and  
$-1\leq\Delta\leq 1$ 
respectively, are indicated by solid circles.}
\label{2den}
\end{figure}
\begin{figure}
    \caption{The sublattice magnetization in the $x$-direction, $M^x$, as a function 
of the anisotropy $\Delta$ for the square-lattice case, from a mean-field calculation and high-order 
ECCM SUB2-n schemes using the $x$-aligned N\'{e}el model state.}
\label{2dxymag}
\end{figure}
\begin{figure}
    \caption{The ground-state energy per spin, $E_g/N$, as a function  
of the orientation of nearest-neighbour spins, $\beta$, in a model state, 
which is restricted to the canted model state 
space, for the linear-chain case, when the relative angle 
$\alpha$ is $\pi$ (N\'{e}el state). The examination of the model-state space
is performed in both the {\it XY}-like and Ising-like regions in the neighbourhood
 of the isotropic point, using the converged ECCM full SUB2 values.}
\label{modst}
\end{figure}
\begin{figure}
   \caption{The ground-state energy per spin, $E_g/N$, as a function  
of the overall orientation of nearest-neighbour spins $\beta$, 
when $\alpha$ is fixed at $\pi$, for the linear-chain case. The examination of the model-state space
is performed in the Ising-like region for increasing values of
the anisotropy, using the converged ECCM full SUB2 values.}
\label{mod}
\end{figure} 
   \epsfxsize=12cm 
   \centerline{\epsffile{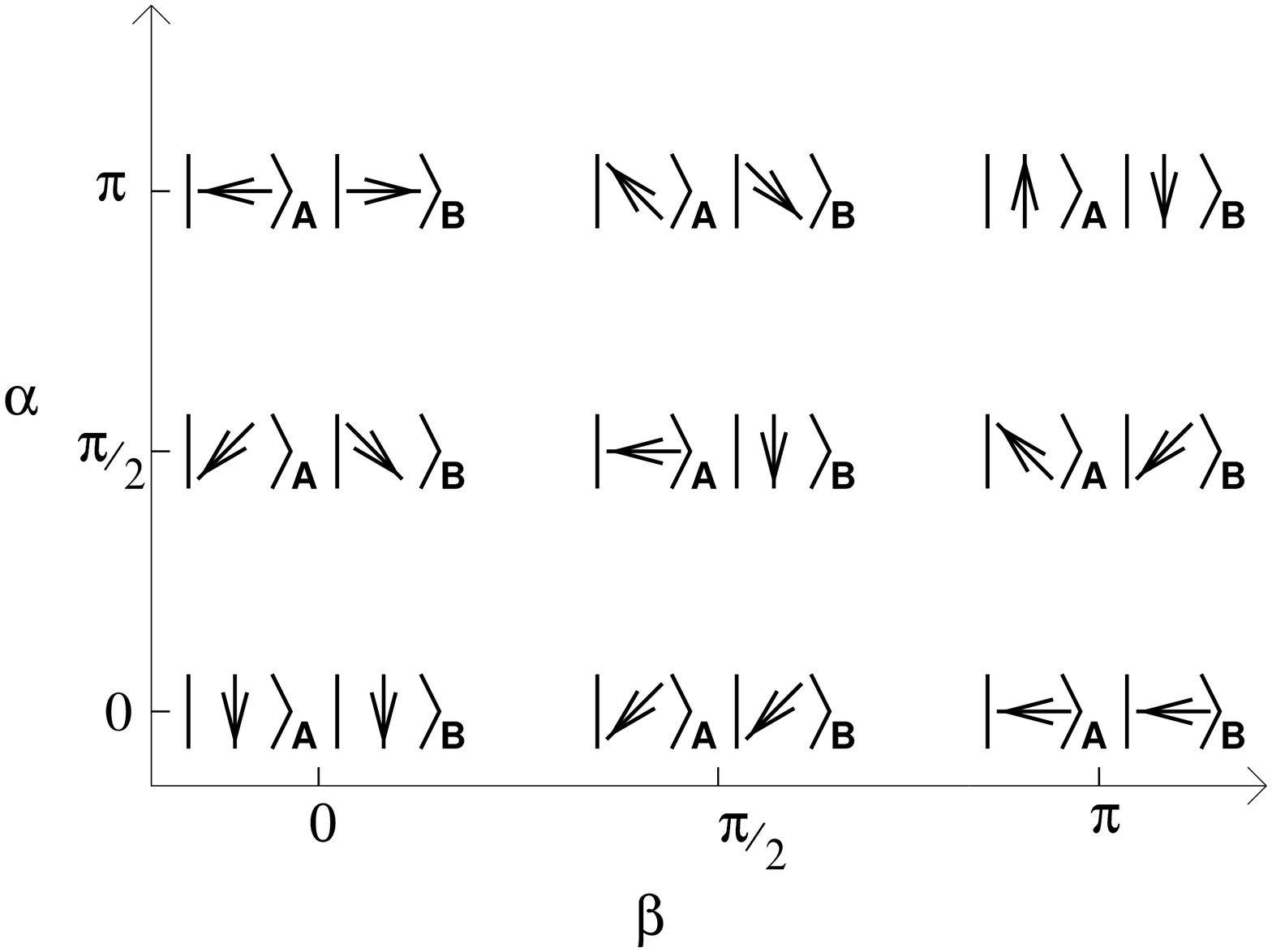}}
\centerline{Fig. 1}
   \epsfxsize=12cm 
   \centerline{\epsffile{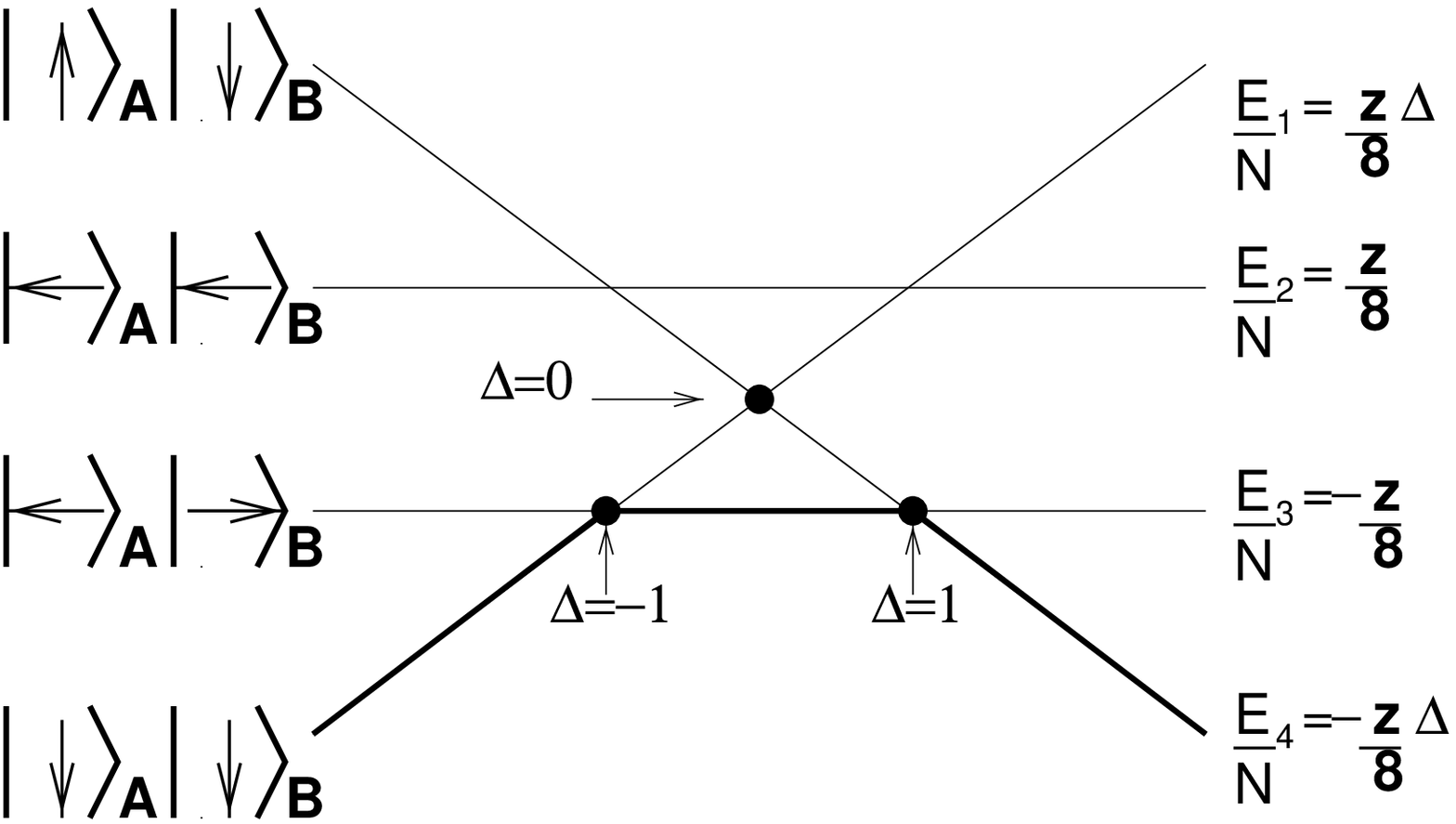}} 
\centerline{Fig. 2}
   \epsfxsize=12cm 
   \centerline{\epsffile{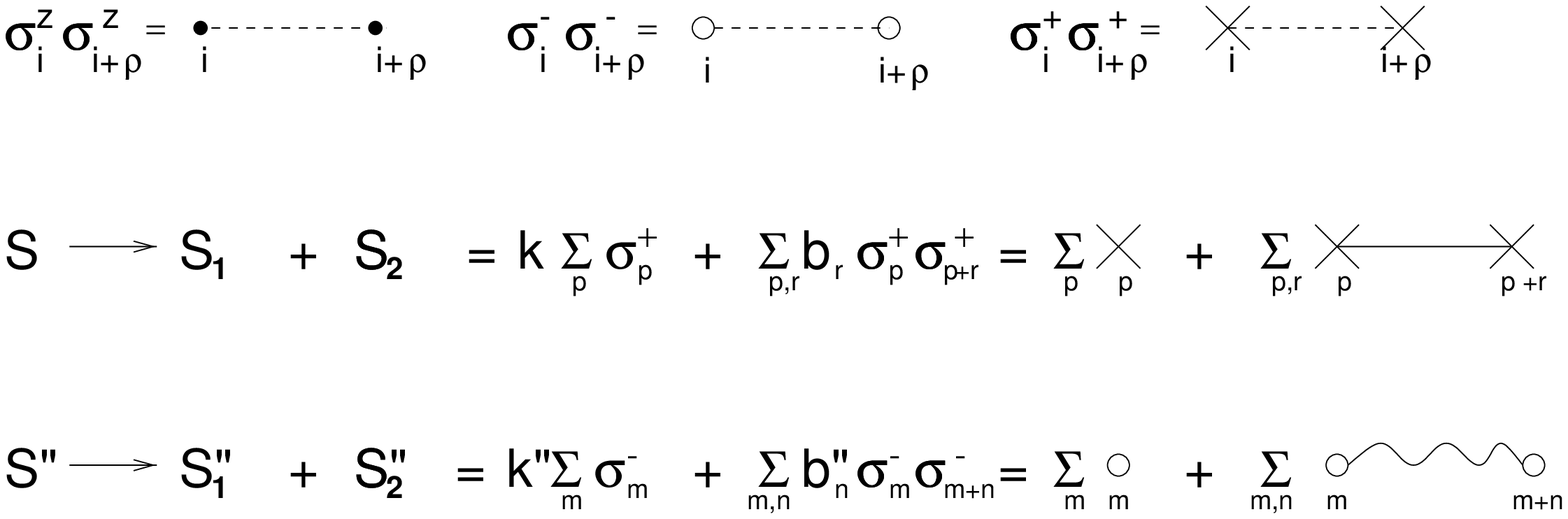}} 
\centerline{Fig. 3}
   \epsfxsize=10cm 
   \centerline{\epsffile{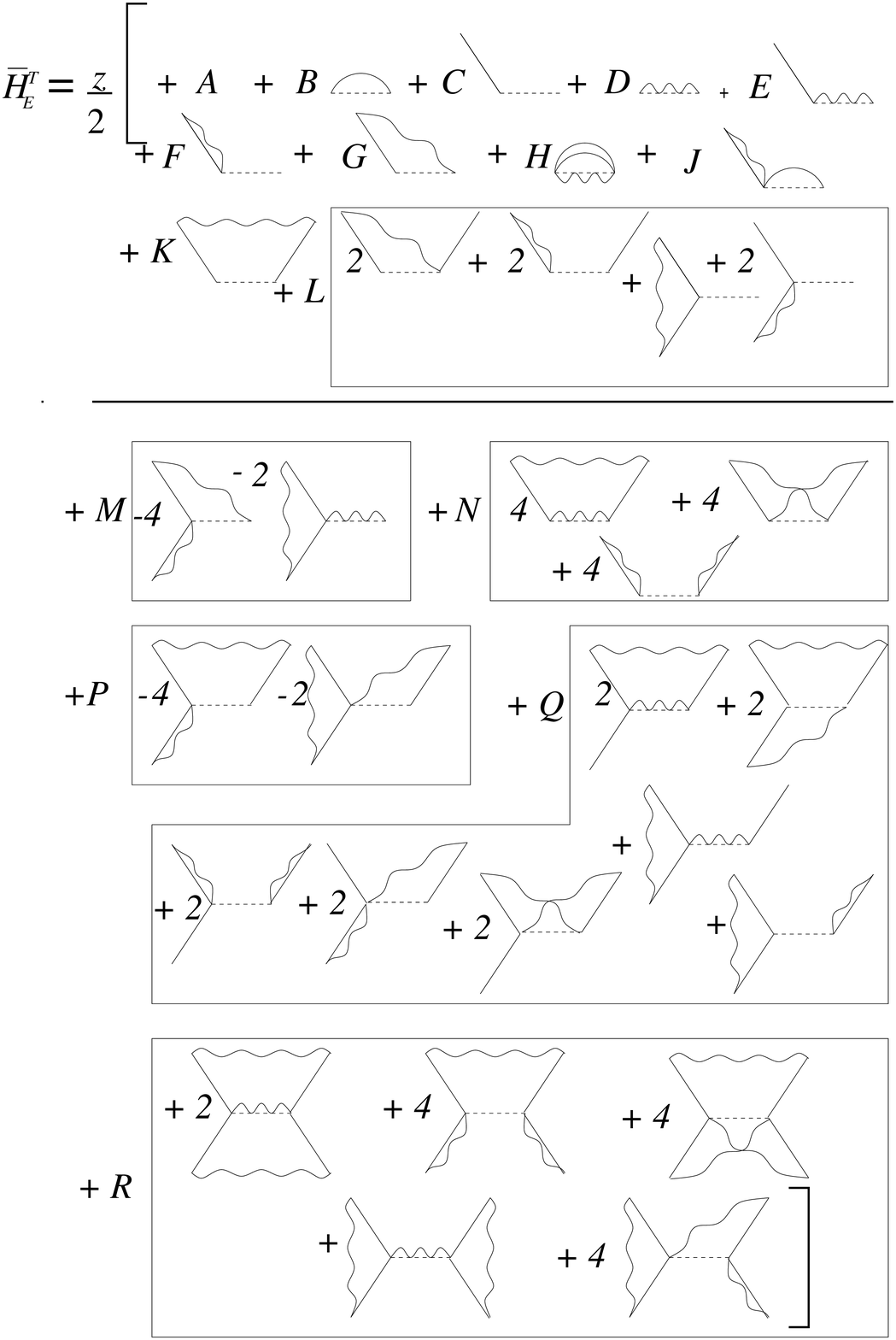}} 
\centerline{Fig. 4}
   \epsfxsize=12cm 
   \centerline{\epsffile{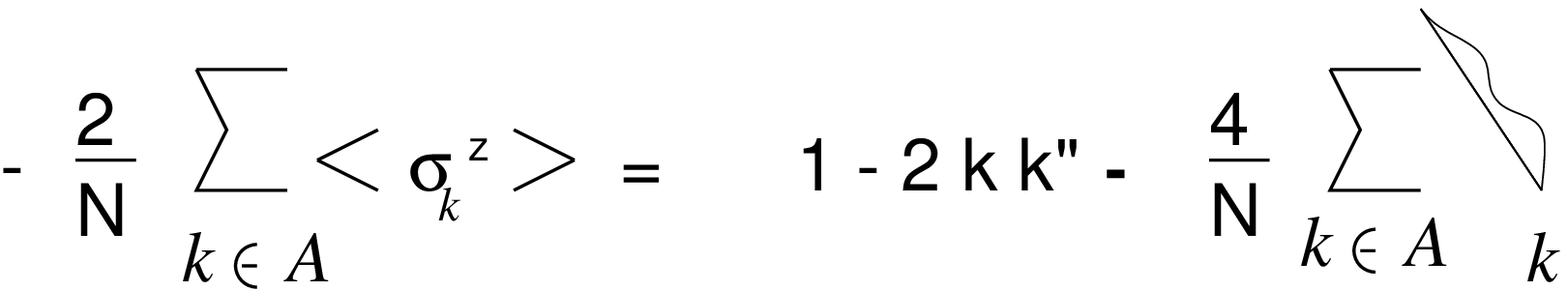}} 
\centerline{Fig. 5}
   \epsfxsize=12cm 
   \centerline{\epsffile{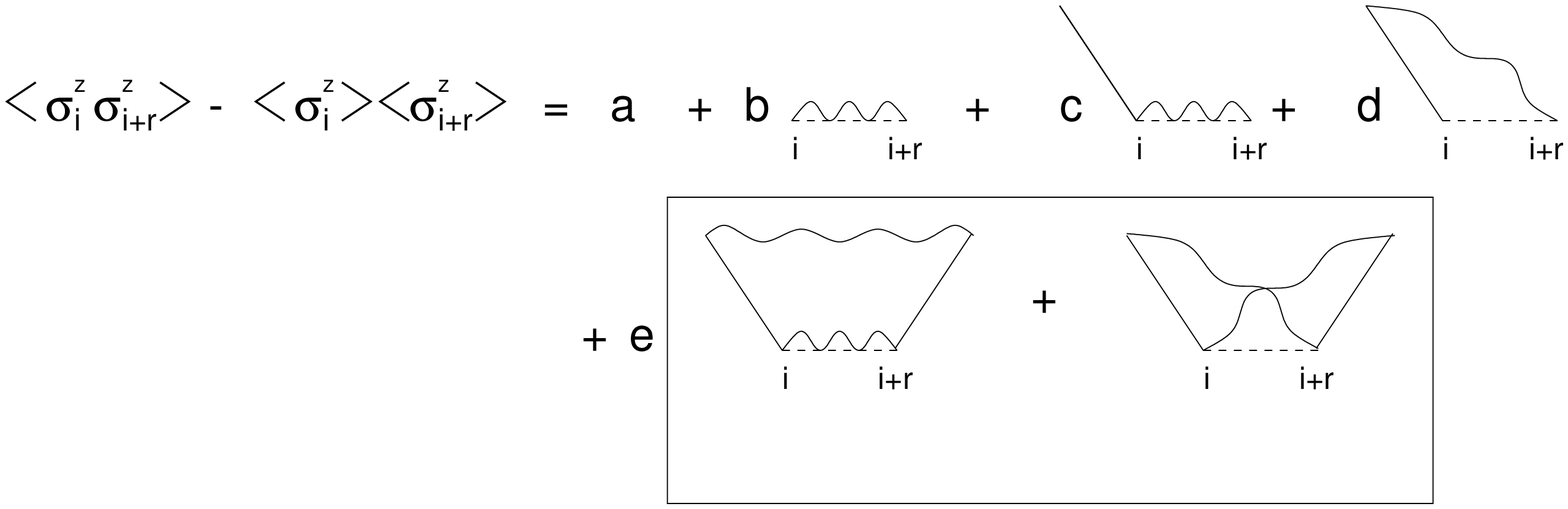}} 
\centerline{Fig. 6}
   \epsfxsize=12cm 
   \centerline{\epsffile{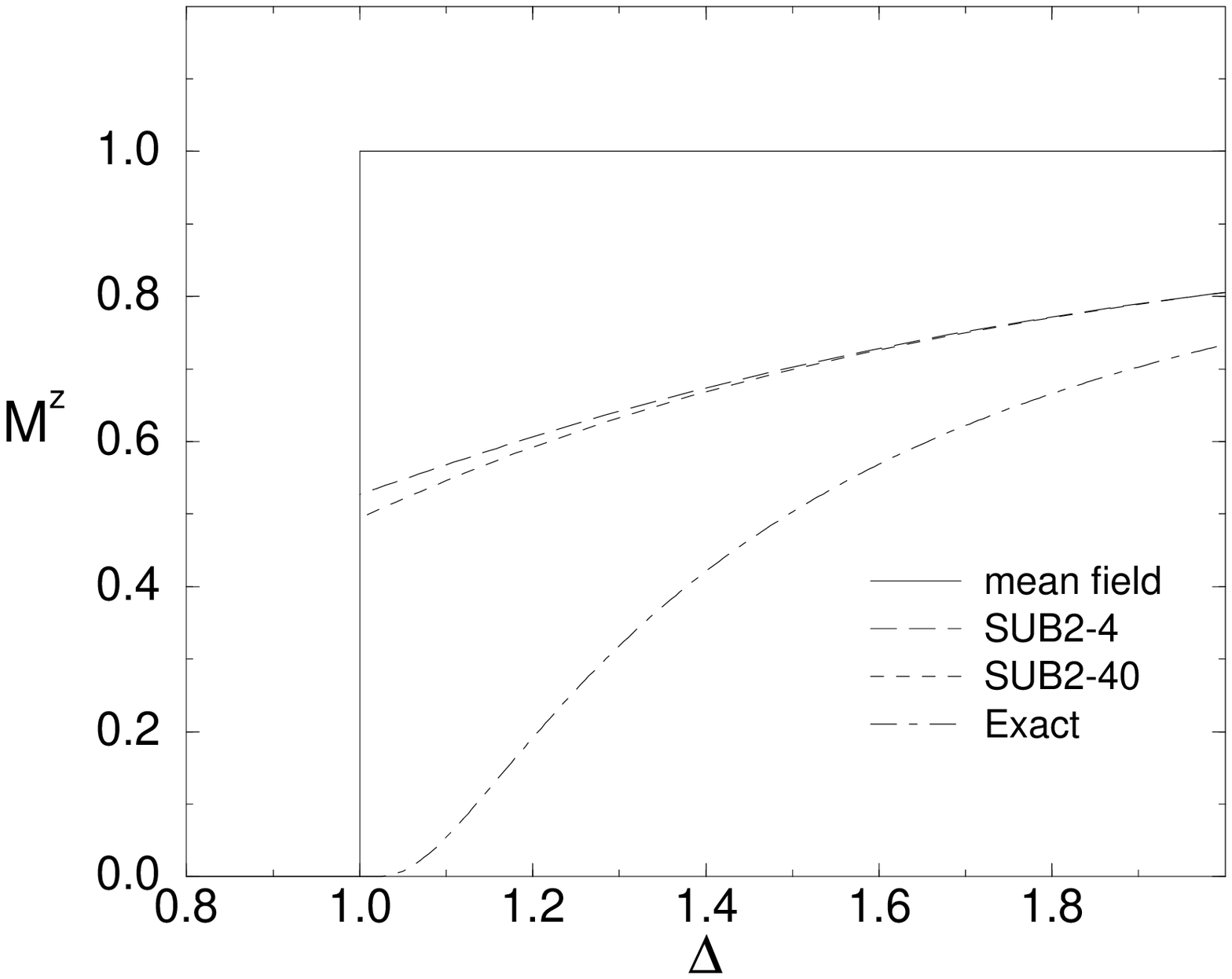}} 
\centerline{Fig. 7}
   \epsfxsize=12cm 
   \centerline{\epsffile{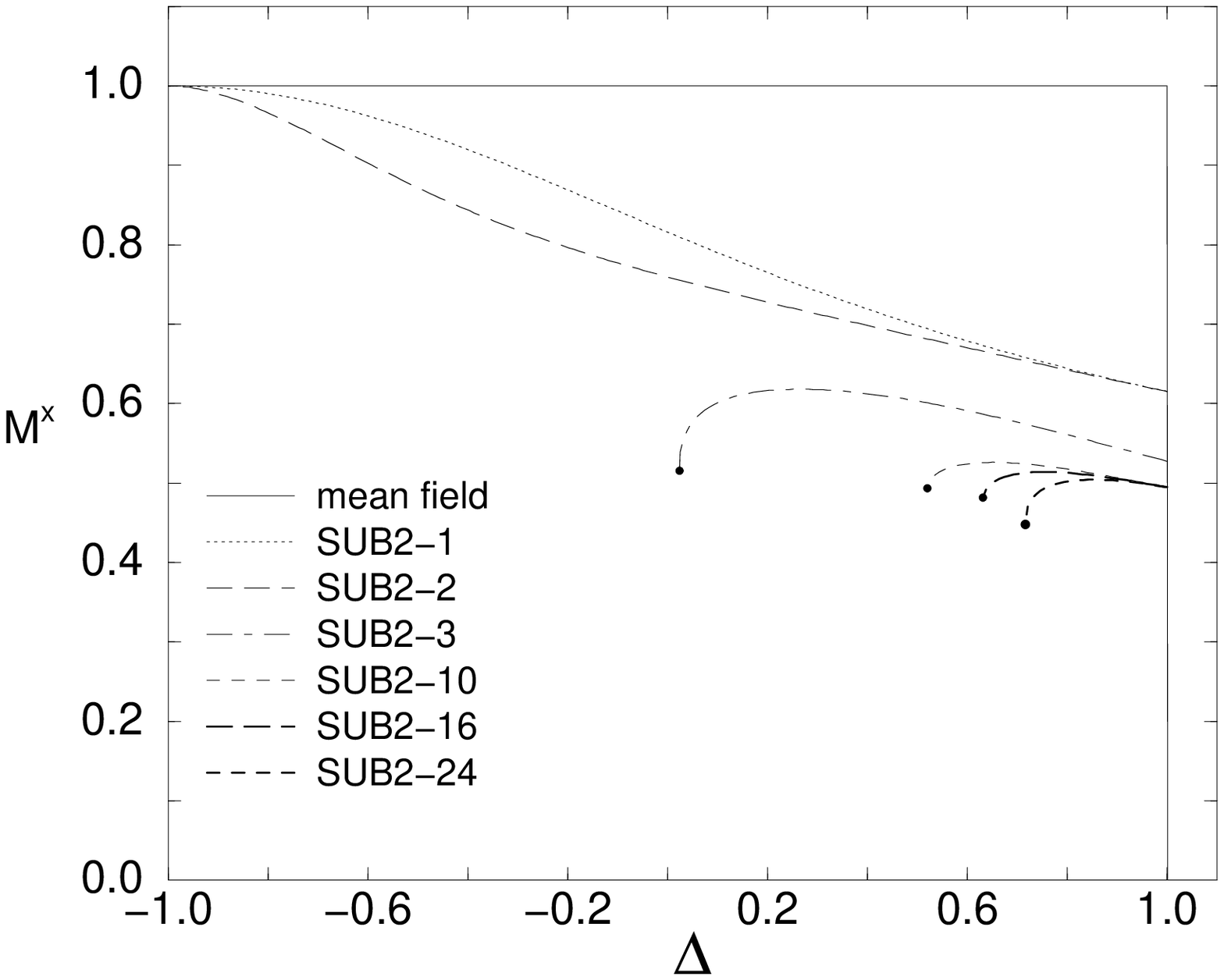}} 
\centerline{Fig. 8}
   \epsfxsize=12cm 
   \centerline{\epsffile{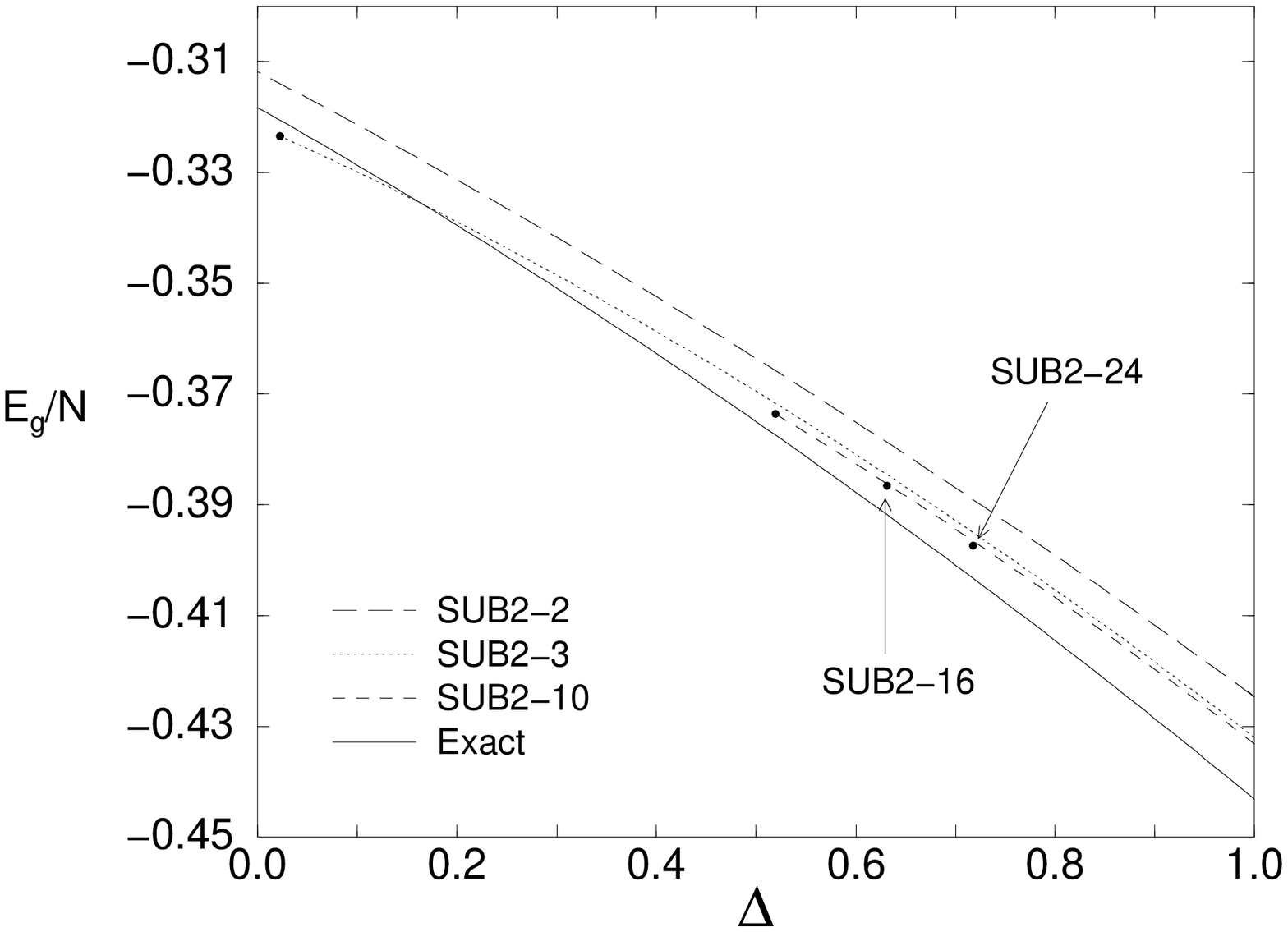}} 
\centerline{Fig. 9}
   \epsfxsize=12cm 
   \centerline{\epsffile{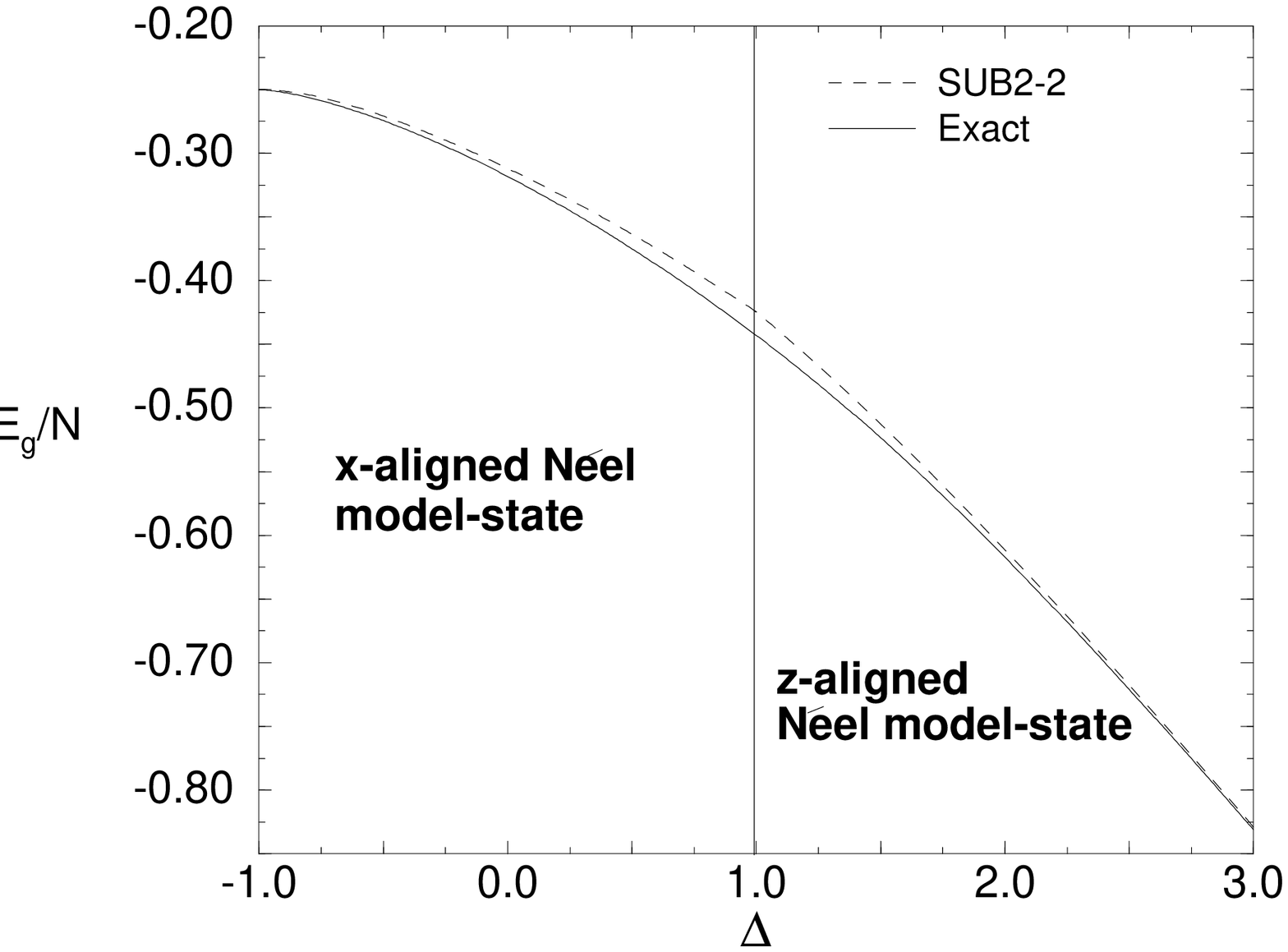}} 
\centerline{Fig. 10}
   \epsfxsize=12cm 
   \centerline{\epsffile{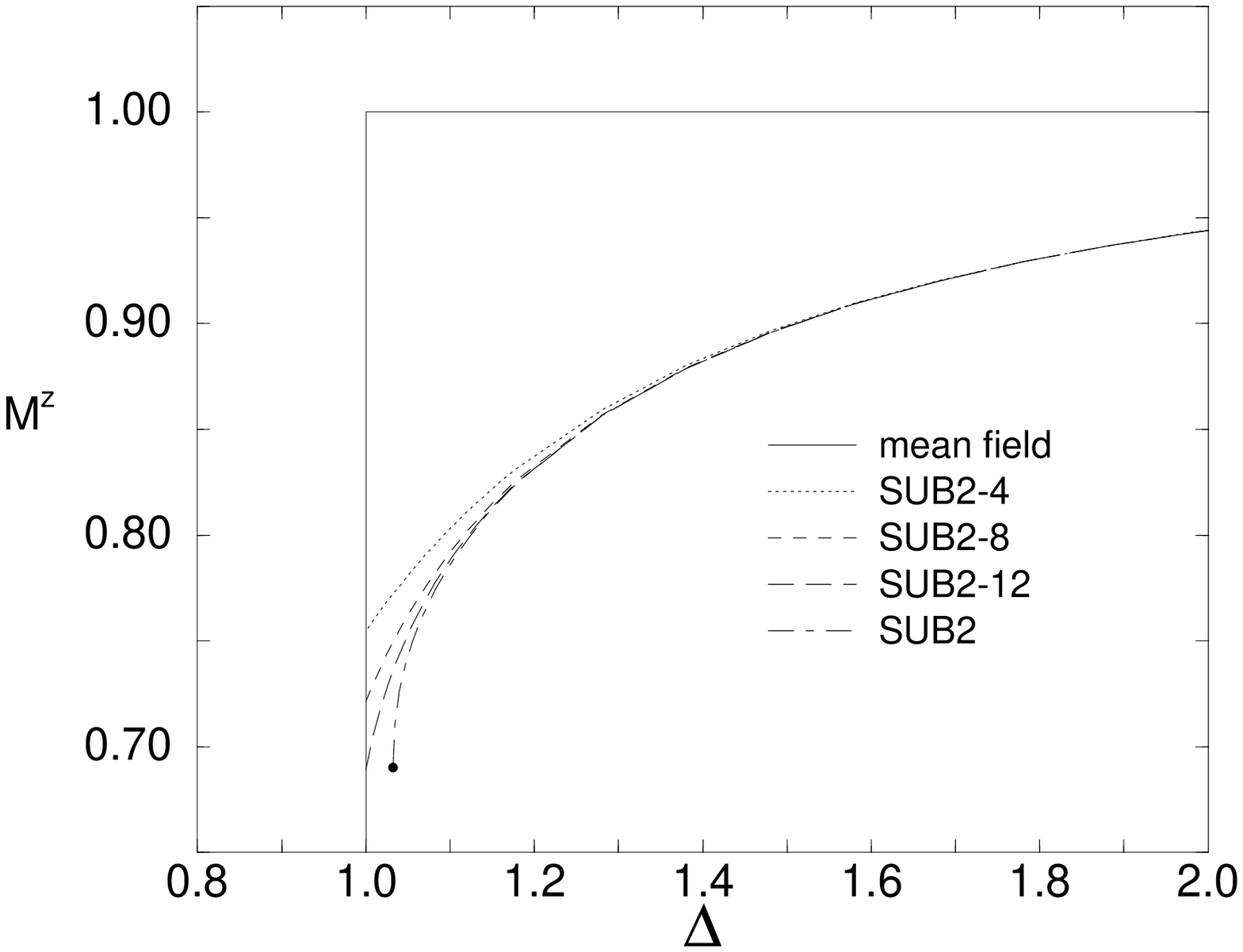}}
\centerline{Fig. 11}
\epsfxsize=12cm 
\centerline{\epsffile{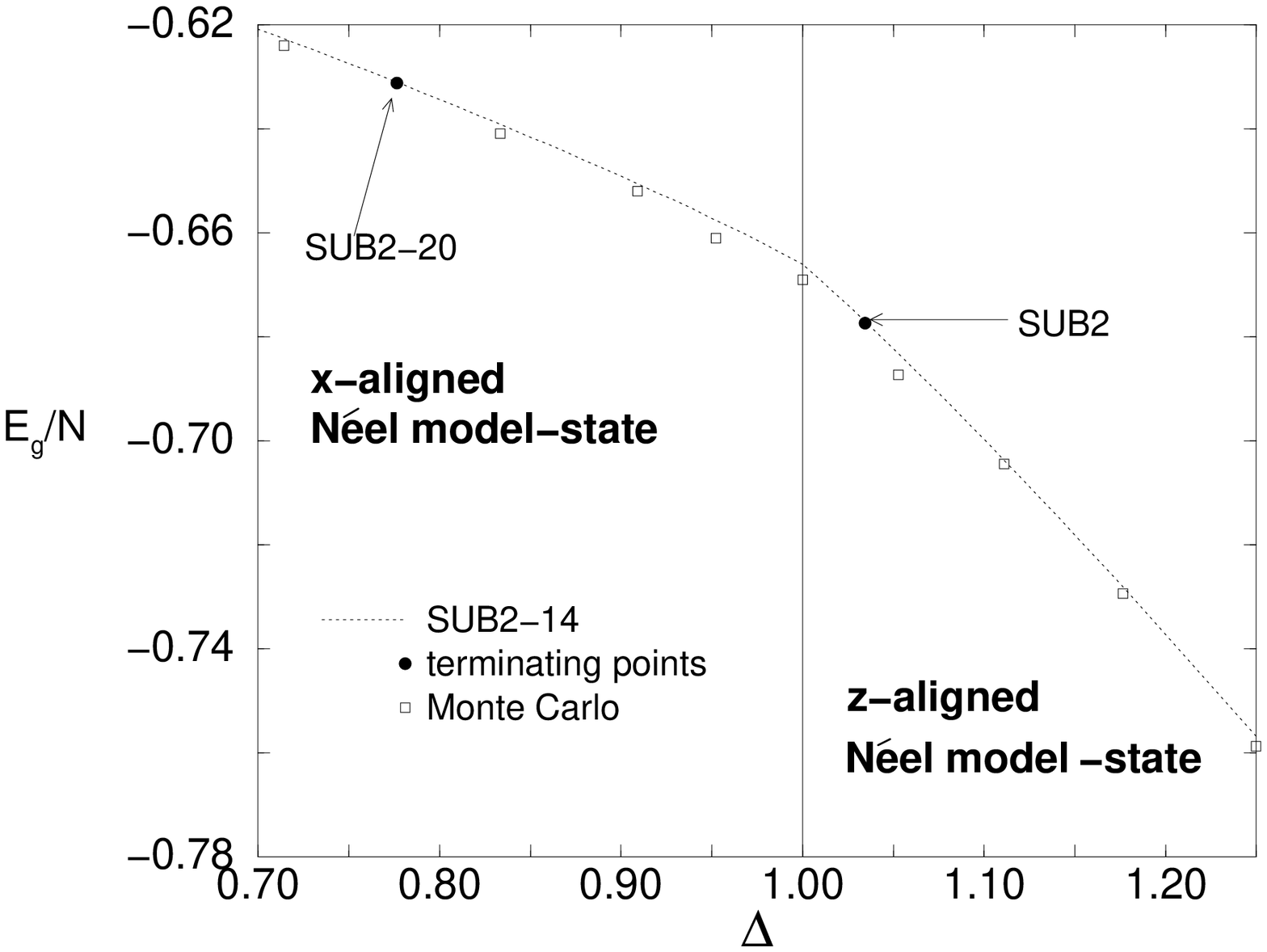}}
\centerline{Fig. 12}
   \epsfxsize=12cm 
   \centerline{\epsffile{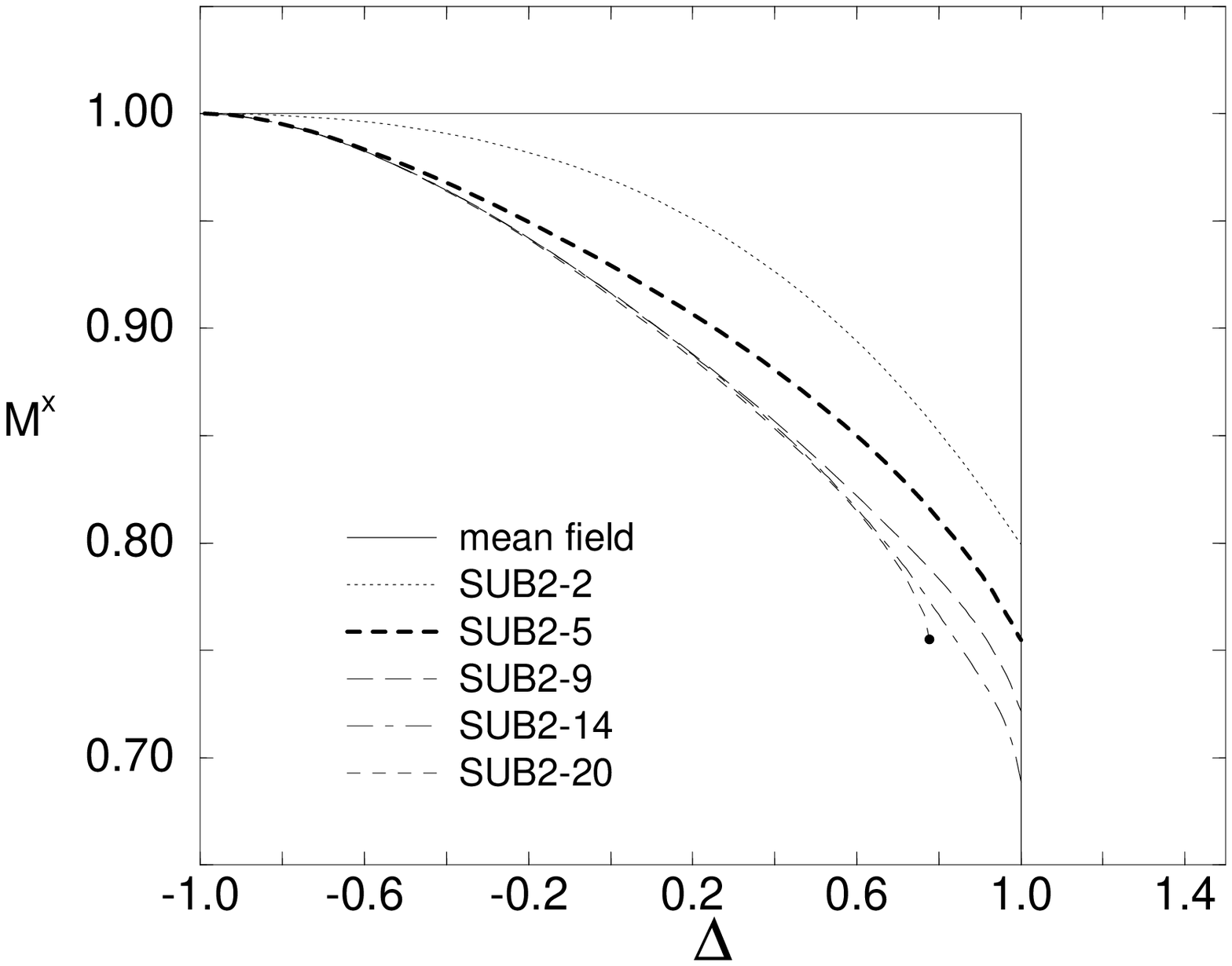}}
\centerline{Fig. 13}
   \epsfxsize=12cm 
   \centerline{\epsffile{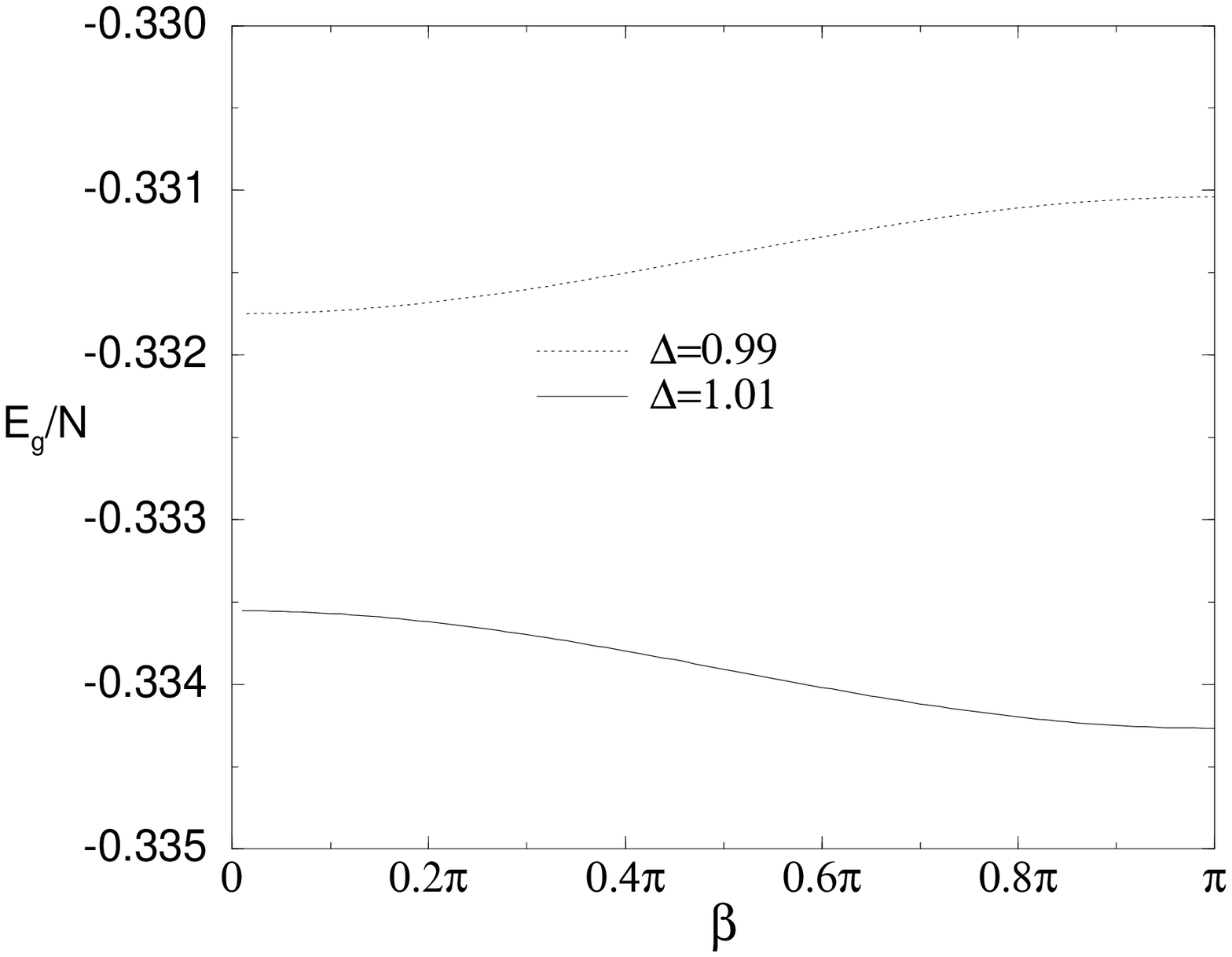}}
\centerline{Fig. 14}
   \epsfxsize=12cm 
   \centerline{\epsffile{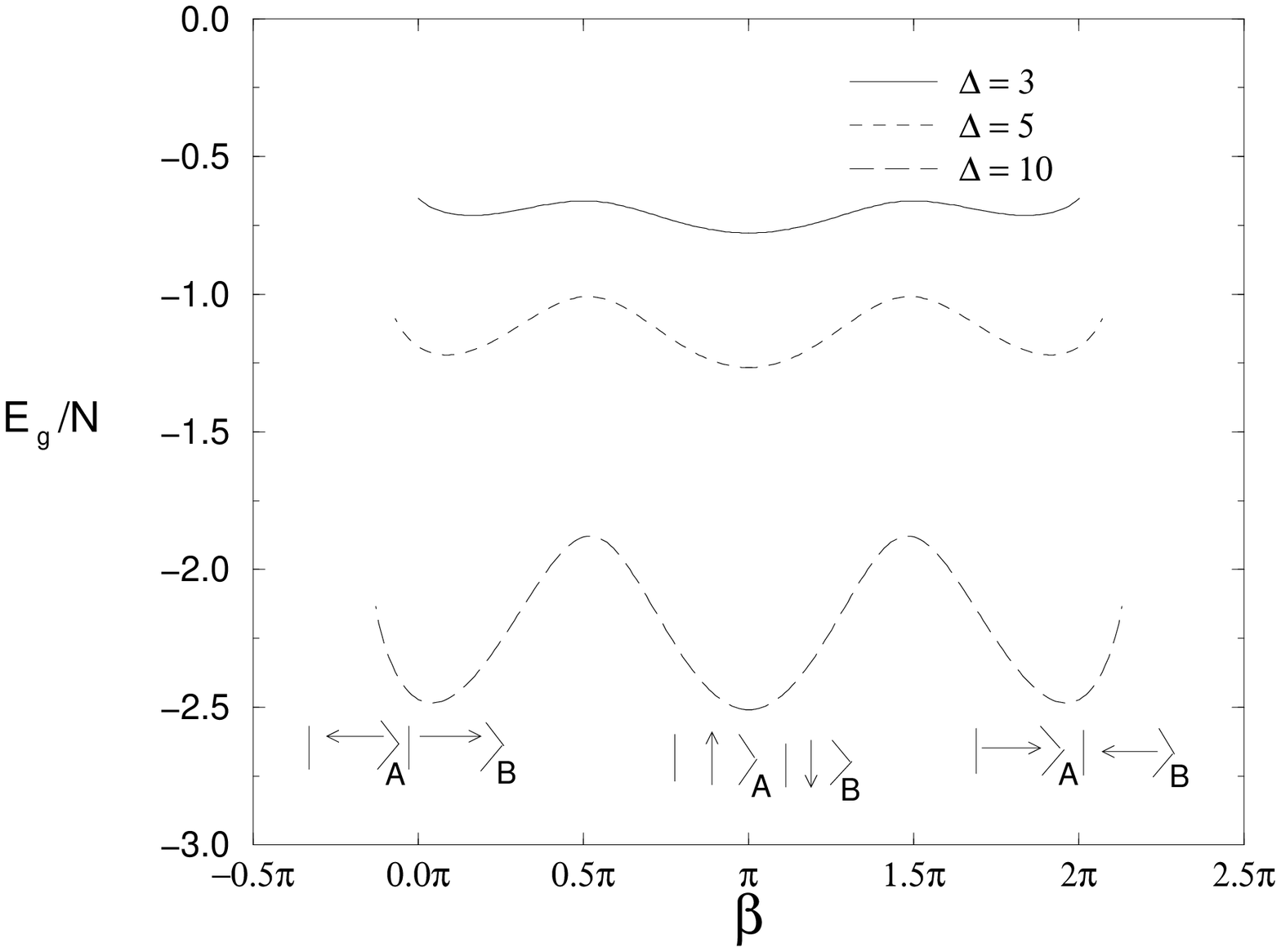}} 
\centerline{Fig. 15}

\end{document}